\documentclass[lettersize,journal]{IEEEtran}
\pdfoutput=1
\usepackage{amsmath,amsfonts}
\usepackage{algorithmic}
\usepackage{algorithm}
\usepackage{array}
\usepackage{textcomp}
\usepackage{stfloats}
\usepackage{url}
\usepackage{verbatim}
\usepackage{graphicx}
\usepackage{cite}
\usepackage{subfigure}
\begin{document}

\title{Terahertz Wireless Data Center: Gaussian Beam or Airy Beam?}

\author{Wenqi Zhao, Sergi Abadal, Guochao Song, Jiamo Jiang, Chong Han,~\IEEEmembership{Senior Member,~IEEE}

        % <-this % stops a space
\thanks{
    Wenqi Zhao is with the Terahertz Wireless Communications (TWC) Laboratory, Shanghai Jiao Tong University, Shanghai 200240,
    China (e-mail: wenqi.zhao@sjtu.edu.cn).
    
    Sergi Abadal is with the NaNoNetworking Center in Catalonia (N3Cat),
Universitat Politècnica de Catalunya, 08034 Barcelona, Spain (e-mail:
abadal@ac.upc.edu).

    Guochao Song and Jiamo Jiang are with China Academy of Information and Communications
Technology, Beijing 100191, China (e-mail: \{songguochao, jiangjiamo\}@
caict.ac.cn).

    Chong Han is with Terahertz Wireless Communications (TWC) Laboratory and also the Cooperative Medianet Innovation Center (CMIC), School of Information Science and Electronic Engineering, Shanghai Jiao Tong University, Shanghai 200240,
    China (e-mail: chong.han@sjtu.edu.cn).

}}

% \markboth{IEEE Transactions on Wireless Communications, Submitted in April 2025}%
% {Shell \MakeLowercase{\textit{Zhao et al.}}: Airy Beam in Terahertz Wireless Data Center}


\maketitle

\begin{abstract}
Terahertz (THz) communication is emerging as a pivotal enabler for 6G and beyond wireless systems owing to its multi-GHz bandwidth. One of its novel applications is in wireless data centers, where it enables ultra-high data rates while enhancing network reconfigurability and scalability.
However, due to numerous racks, supporting walls, and densely deployed antennas, the line-of-sight (LoS) path in data centers is often instead of fully obstructed, resulting in quasi-LoS propagation and degradation of spectral efficiency.
To address this issue, Airy beam-based hybrid beamforming is investigated in this paper as a promising technique to mitigate quasi-LoS propagation and enhance spectral efficiency in THz wireless data centers. Specifically, a cascaded geometrical and wave-based channel model (CGWCM) is proposed for quasi-LoS scenarios, which accounts for diffraction effects while being more simplified than conventional wave-based model. Then, the characteristics and generation of the Airy beam are analyzed, and beam search methods for quasi-LoS scenarios are proposed, including hierarchical focusing-Airy beam search, and low-complexity beam search. Simulation results validate the effectiveness of the CGWCM and demonstrate the superiority of the Airy beam over Gaussian beams in mitigating blockages, verifying its potential for practical THz wireless communication in data centers.

\end{abstract}


\section{Introduction}
\IEEEPARstart{O}{wing} to the demand for ultra-broad multi-GHz bandwidth, Terahertz (THz) wireless communications are envisioned as an enabling technology for six generation (6G) and beyond systems~\cite{Akyildiz-2022-Terahertz}. Among the various promising applications, one notable use case is Tbps-scale Internet of Thing (IoT) connectivity in wireless data centers, which can not only support the high transmission rate with ultra-wideband but also enhances the reconfigurability and scalability of the data center~\cite{Akyildiz-2022-Terahertz,Chen-2019-A-survey,Hamza-2016-Wireless}. In such environments, where communication distances are typically short, and the channel conditions are quasi-static, THz wireless links are particularly suitable for high speed, low-latency data transmission. Furthermore, in wireless data centers,  ultra-massive MIMO (UM-MIMO) with large array apertures can provide near-field multiplexing gains~\cite{An-2024-Near-Field}, while hybrid beamforming techniques help mitigate severe path loss and enhance spectral efficiency \cite{Han-2021-Hybrid}.
 Therefore, the application of THz technology in wireless data centers holds great promise for the future.

\begin{figure}[t]
    \centering
    \includegraphics[width=\linewidth]{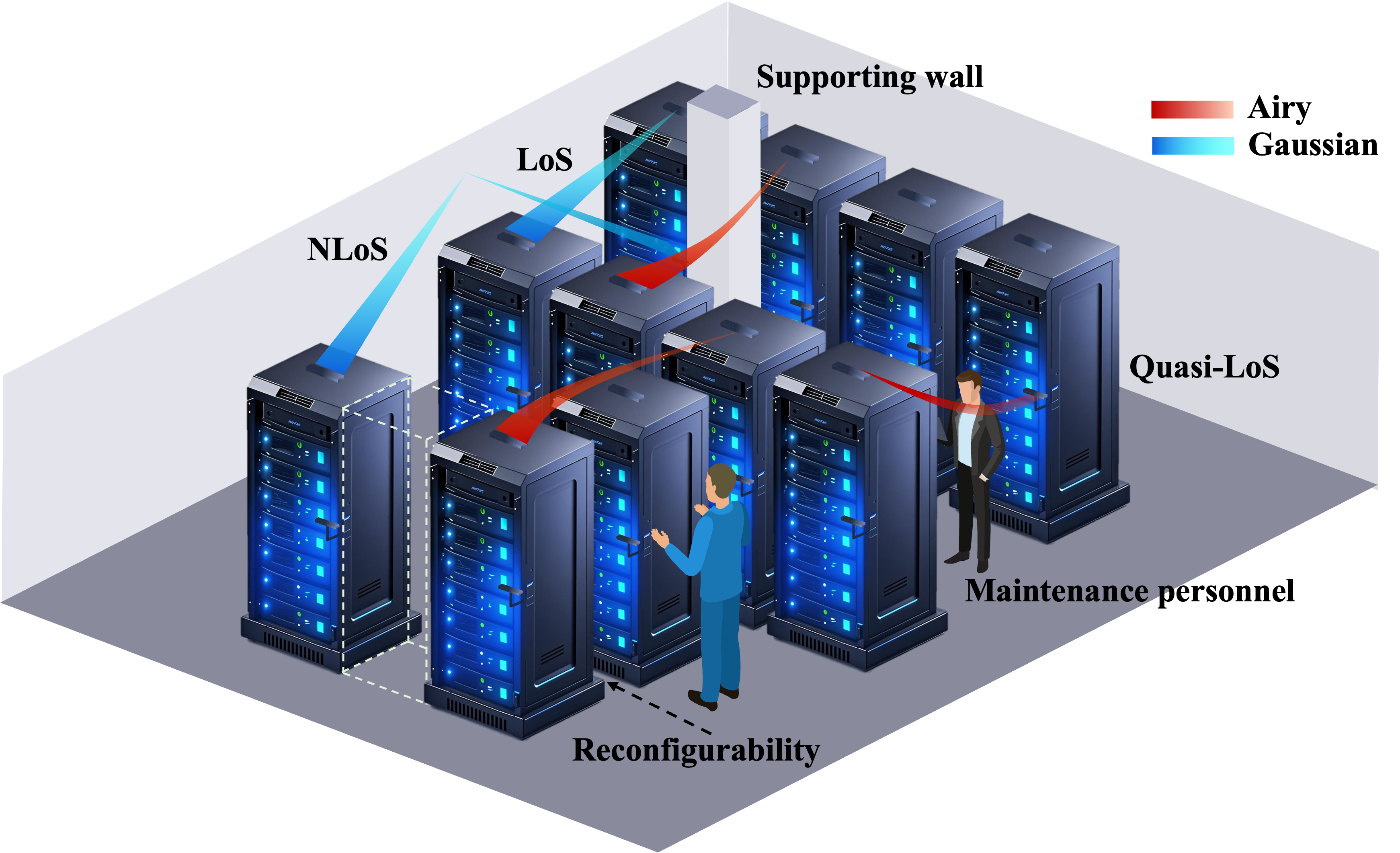} % 使用 \linewidth 自适应栏宽
    \caption{THz wireless data center with LoS, quasi-LoS and NLoS.}
     \label{fig:data_center}
\end{figure}

However, as shown in Fig.~\ref{fig:data_center}, the environment in a data center is complex, with numerous racks, supporting walls, and densely arranged antenna arrays. During the reconfiguration, scaling, and maintenance of the data center, communication links can be easily obstructed by antenna arrays, racks, servers, and personnel, leading to undesirable interruptions and reduced spectral efficiency. These obstructions often lead to partial or complete blockage of the LoS, resulting in quasi-LoS or NLoS conditions.
Currently, state-of-the-art hybrid beamforming algorithms are typically designed for scenarios with either LoS or NLoS since the partial blockage situations are often negligible in long-distance communication. With the increase in the antenna aperture size and the reduction of the communication distance, quasi-LoS becomes more common, especially in wireless data centers, which are more crowded and dense with racks and antenna arrays compared to outdoor free-space environments \cite{Cheng-2020-Characterization}. To this end, beamforming design and beam search schemes for quasi-LoS scenarios are necessary and significantly needed in the THz UM-MIMO systems within the wireless data centers.

\subsection{Related Works}
\subsubsection{Wireless Data Center} The wireless data center has been proposed in the past two decades \cite{Cui-2018-Wireless,Hamza-2016-Wireless,Mamun-2018-Performance, Rommel-2018-Data, Ahearne-2019-Integrating} to realize fast deployment and dynamic reconfigurability. In \cite{Hamza-2016-Wireless,Mamun-2018-Performance}, free space optical (FSO) and 60 GHz radio frequency are proposed to be the two key candidate technologies for wireless link in data centers to realize the top-of-rack (ToR)-to-ToR wireless communication. Additionally, with the recent advances in THz wireless technologies, the feasibility of integrating wireless THz links into data center network is investigated \cite{Rommel-2018-Data, Ahearne-2019-Integrating} to enhance the reconfigurability and maintain a high transmission rate due to ultra-high bandwidth. 
Recently, some channel measurements in the THz band have been conducted \cite{Channel-2022-Song,Cheng-2020-Characterization,Cheng-2020-THz,Eckhardt-2024-Channel}.
The measurement campaigns at the 130–140 GHz band \cite{Channel-2022-Song} have shown that reflections and scattering from metal racks significantly impact channel characteristics, leading to lower path loss, smaller K-factor, and larger delay spreads compared to conventional indoor scenarios. In \cite{Cheng-2020-Characterization,Cheng-2020-THz}, the authors have conducted channel measurements at 300 GHz, analyzing rack-to-rack and blade-to-blade communications while evaluating LoS, NLoS, and quasi-LoS links. A recent study \cite{Eckhardt-2024-Channel} conducted double-directional channel measurements at 300~GHz, also analyzing these three propagation scenarios, which focused on channel parameters like path loss and delay spread, revealing the impact of scattering and multi-path propagation, particularly for high signal-to-noise ratio systems.

These works demonstrate that THz communication holds significant potential for enabling wireless connectivity in data centers. However, most of these efforts have primarily focused on channel measurement and modeling, while practical beamforming strategies and beam search schemes under quasi-LoS conditions need further study.
%Gaussian beam in this paper indicate to the far field beam steering and near field beam focusing this no-diffraction beam
\subsubsection{Gaussian Beam and Airy Beam}
From an electromagnetic perspective, a Gaussian beam exhibits a transverse power distribution that follows a Gaussian function which is commonly employed for beam steering in the far field~\cite{Alkhateeb-2015-Limited} and beam focusing in the near field~\cite{Zhang-2022-Beam-focusing, Cui-2022-Channel-estimation, Cui-2023-Near-Field-MIMO, Wu-2023-Multiple}. These beams exhibit good focusing and diffraction characteristics, but their energy tends to diverge as the propagation distance increases. In contrast, the Airy beam, first proposed in the field of optics \cite{Zhan-2020-Propagations, Nikolaos-2019-Airy}, features a transverse field distribution following the Airy function, which results its unique properties such as non-diffraction, self-acceleration, and self-healing. Recently, the Airy beam has been proposed for use in the THz band \cite{Jornet-2023-Wireless, Petrov-2024-Wavefront}. One notable application of the Airy beam is its ability to curve around obstacles \cite{Chen-2024-Curving, Guerboukha-2024-Curving, Petrov-2024-Wavefront-Hopping, Singh-2024-Wavefront} due to its self-acceleration feature, enabling a curved trajectory that allows it to have a good performance in quasi-LoS scenarios. In \cite{Guerboukha-2024-Curving}, the authors conduct experimental measurements demonstrating that Airy beams carrying high-data-rate transmissions can establish links by bending around obstacles. In \cite{Chen-2024-Curving}, the authors investigate the use of Airy beams in a THz uniform linear array (ULA) antenna array and propose a physics-informed learning-based framework to optimize the phase profile of the transmitting array, enabling the resulting wavefront to curve around obstacles. Simulation results indicate that the Airy beam outperforms conventional beam steering and beam focusing. However, this method requires prior knowledge of the environment, which is also difficult and uses much overhead. 

Although these studies have suggested the potential of Airy beams for bypassing blockages 
 in quasi-LoS scenarios, they do not provide a detailed implementation framework for THz UM-MIMO systems. Given that quasi-LoS propagation is prevalent in THz wireless data centers, it is essential to investigate practical Airy beam implementations and develop efficient beam search schemes to ensure reliable link establishment in such environments.
\subsection{Contributions}
To address the aforementioned challenges, in this paper, we investigate Airy beam-based beamforming in a quasi-LoS environments for THz wireless data centers. We propose a novel cascaded geometric and wave channel model (CGWCM) to capture the diffraction characteristics in the blocked scenarios effectively. Furthermore, three Airy beam search schemes are developed to establish communication links under quasi-LoS. Extensive simulation results demonstrate that the proposed Airy beam search schemes outperform conventional Gaussian beam search approaches in quasi-LoS environments, offering a more practical solution for THz wireless data centers. The main contributions are summarized as follows.
\begin{itemize}
\item[$\bullet$] \textbf{We develop a CGWCM for the quasi-LoS scenarios, which achieves a balance between accuracy and complexity.} Specifically, we consider the ULA in the transceiver and present the geometrical channel model (GCM) and the wave-based channel model (WCM). In quasi-LoS, GCM exhibits low accuracy because it neglects diffraction, whereas the WCM can accurately capture the diffraction effects caused by the blockages but involves a complex integral form.
To establish an accurate and low-complexity model, we develop the CGWCM, incorporating the concept of virtual arrays in blocked regions, integrating GCM and WCM, and transforming the complex integral form into a simple matrix product form. Finally, the calibration of the CGWCM is discussed to enable practical spectral efficiency evaluation in quasi-LoS scenarios.
% 说清楚各个
%我们研究了如何基于THz hybrid beamforming生成Airy beam的方式同时分析了其三个特性
\item[$\bullet$] \textbf{Motivated by the non-diffraction, self-acceleration, and self-healing characteristics, we investigate the Airy beam-based beamforming design.} Three key characteristics, non-diffraction, self-acceleration, and self-healing, are introduced, which are helpful for the transmission of the Airy beams in the blocked scenarios. We further explore the analog precoder of Airy beams by imposing a cubic phase profile on a Gaussian beam, which can be practically implemented by adjusting the phase shifts in THz hybrid beamforming architectures. Then, the digital precoder in the Tx and combiner in the Rx are designed with the Airy beam-based analog precoder and the available channel information.

\item[$\bullet$] \textbf{We propose Airy beam search schemes for the communication link establishment in the quasi-LoS scenarios without prior knowledge of blockage position.} We first derive the sampling interval by analyzing the correlation between arbitrary Airy beams and obtain an exhaustive search scheme for optimal beam with high training overhead. To reduce this overhead, a hierarchical focusing-Airy beam search is proposed, which achieves near-optimal beam by first identifies the best focusing beam and then refining the curving coefficient. To further reduce overhead and accelerate link establishment, a low-complexity beam search is proposed, which uses prior position information to steer the beam toward the receiver and obtain a sub-optimal Airy beam.
% we use measured data (用了实测数据)
\item[$\bullet$] \textbf{We evaluate the performance of the CGWCM and proposed beam search schemes based on real channel data measured in a data center.} Results demonstrate that the CGWCM remains more accurate than GCM in quasi-LoS scenarios while having a more simplified form than WCM. The proposed hierarchical focusing-Airy beam search scheme outperforms the Gaussian beam search schemes and achieves performance comparable to exhaustive search schemes but with reduced overhead. Meanwhile, the proposed low-complexity beam search scheme can rapidly establish communication links while maintaining good performance with significantly lower overhead.
\end{itemize}

%，提出了简化的波动模型，对衍射的波动模型与几何模型进行了结合
\subsection{Organization}
The remainder of the paper is organized as follows. In Sec.~\ref{sec:System Model}, the system model of THz UM-MIMO systems is provided. In Sec.~\ref{sec:CGWCM}, we introduce the GCM and WGM. By combining these two models, the low-complexity CGWCM model is proposed. Then, we analyze the characteristics and  Airy beam-based beamforming design in Sec.~\ref{Sec:Analysis of Airy}. Furthermore, in Sec.~\ref{sec:beam search}, the beam correlation is analyzed and beam search schemes are proposed. After an in-depth analysis and numerical evaluation of the proposed CGWCM and beam search schemes in Sec.~\ref{sec: evaluation}, the paper is summarized in Sec.~\ref{sec:conclusion}.

\section{System Model}
\label{sec:System Model}
In this section, we introduce the hybrid beamforming architecture in a wireless data center.
% Tx and Rx system model
We consider the hybrid beamforming architecture with ULA at both the Tx and Rx, respectively. To transmit $N_s$ data streams, there are $N_t$ transmit antennas and $N_r$ receive antennas with $L_t$ and $L_r$ RF chains in the Tx and Rx, which satisfy $N_s\leq L_t < N_t$ and $N_s \leq L_r < N_r$. For analysis simplicity, $L_t=L_r=N_s$ is assumed throughout the paper. In the Tx, the signal sequentially passes through the digital precoder $\mathbf{F}_{\mathrm{BB}}\in\mathbb{C}^{L_t\times N_s}$ and analog precoder $
\mathbf{F}_{\mathrm{RF}}\in\mathbb{C}^{N_t\times L_t}$ where each element satisfies the constant module constraint, i.e., $|\mathbf{F}_{\mathrm{RF}}(i,j)|=\frac{1}{\sqrt{N_t}}$. Additionally, $\mathbf{F}_{\mathrm{BB}}$ should be normalized to satisfy $||\mathbf{F}_{\mathrm{RF}}\mathbf{F}_{\mathrm{BB}}||_\mathrm{F}^2=N_s$ to meet the total transmit power constraint. After transmitted to the channel $\mathbf{H}\in\mathbb{C}^{N_r \times N_t}$, the signal is received by an analog combiner $\mathbf{W}_\mathrm{RF}\in\mathbb{C}^{N_r\times L_r}$and a digital combiner $\mathbf{W}_\mathrm{BB}\in\mathbb{C}^{L_r\times N_s}$. Similar to the Tx, the analog combiner also follows the constant module constraint, i.e., $|\mathbf{W}_{\mathrm{RF}}(i,j)|=\frac{1}{\sqrt{N_r}}$ and the digital combiner should be normalized to satisfy $||\mathbf{W}_{\mathrm{BB}}^\mathrm{H}\mathbf{W}_{\mathrm{RF}}^\mathrm{H}||^2_{\mathrm{F}}=N_s$. The received signal can be represented as
\begin{equation}
    \mathbf{y}=\sqrt{\rho}\mathbf{W}_\mathrm{BB}^\mathrm{H}\mathbf{W}_\mathrm{RF}^\mathrm{H}\mathbf{H}\mathbf{F}_\mathrm{RF}\mathbf{F}_\mathrm{BB}\mathbf{s}+\mathbf{W}_\mathrm{BB}^\mathrm{H}\mathbf{W}_\mathrm{RF}^\mathrm{H}\mathbf{n},
\end{equation}
where $\mathbf{s}$ and $\mathbf{y}$ are $N_s\times 1$ transmitted and received signals. $\rho$ is the transmit power and $\mathbf{n} \sim \mathcal{CN}(0,\sigma^2\mathbf{I}_{N_r})$ represents the circularly symmetric complex Gaussian distributed additive noise vector, with $\sigma^2$ noise power.
% \begin{figure}[t]

%     \centering
%     \includegraphics[width=0.8\linewidth]{images/system_model/system_model.png} % 使用 \linewidth 自适应栏宽
%     \caption{Hybrid beamforming architecture of UM-MIMO communication systems.}
%      \label{fig:system_model}
% \end{figure}
\section{Cascaded Geometric and Wave Channel Model}
\label{sec:CGWCM}
In this section, we first introduce the GCM and WCM. Then, the CGWCM is investigated, which accounts for diffraction effects while owning lower complexity than the WCM. Finally, the calibration of the three models is discussed to enable practical spectral efficiency evaluation in quasi-LoS scenarios.

%每个subsection的最后加一个总结，内容可以去看contribution的部分
\subsection{GCM}
\begin{figure*}[ht]
    \centering
    \subfigure[Geometrical channel model.]{\includegraphics[width=0.32\linewidth]{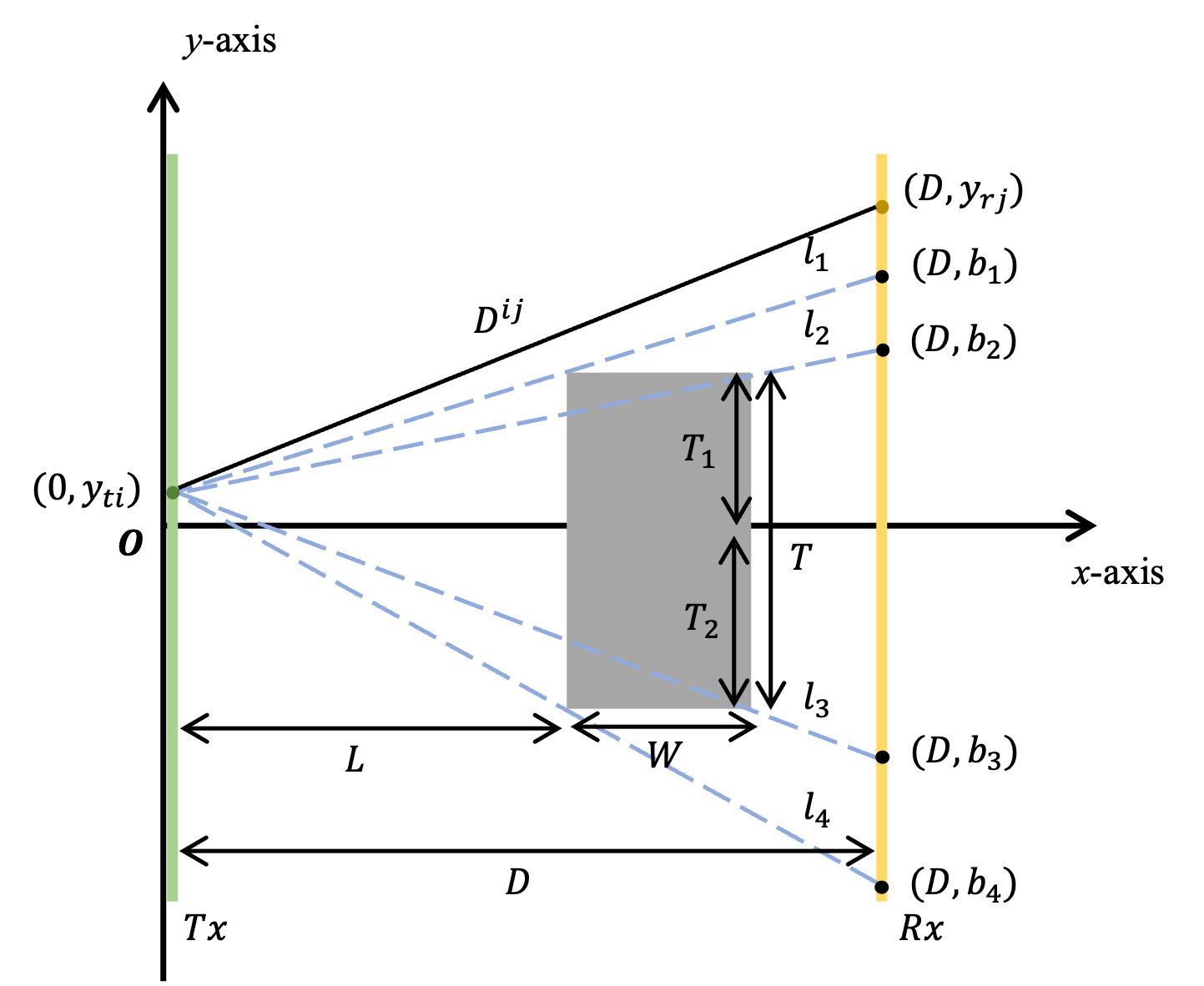}\label{fig:GCM}} 
     \subfigure[Wave-based channel model.]{\includegraphics[width=0.33\linewidth]{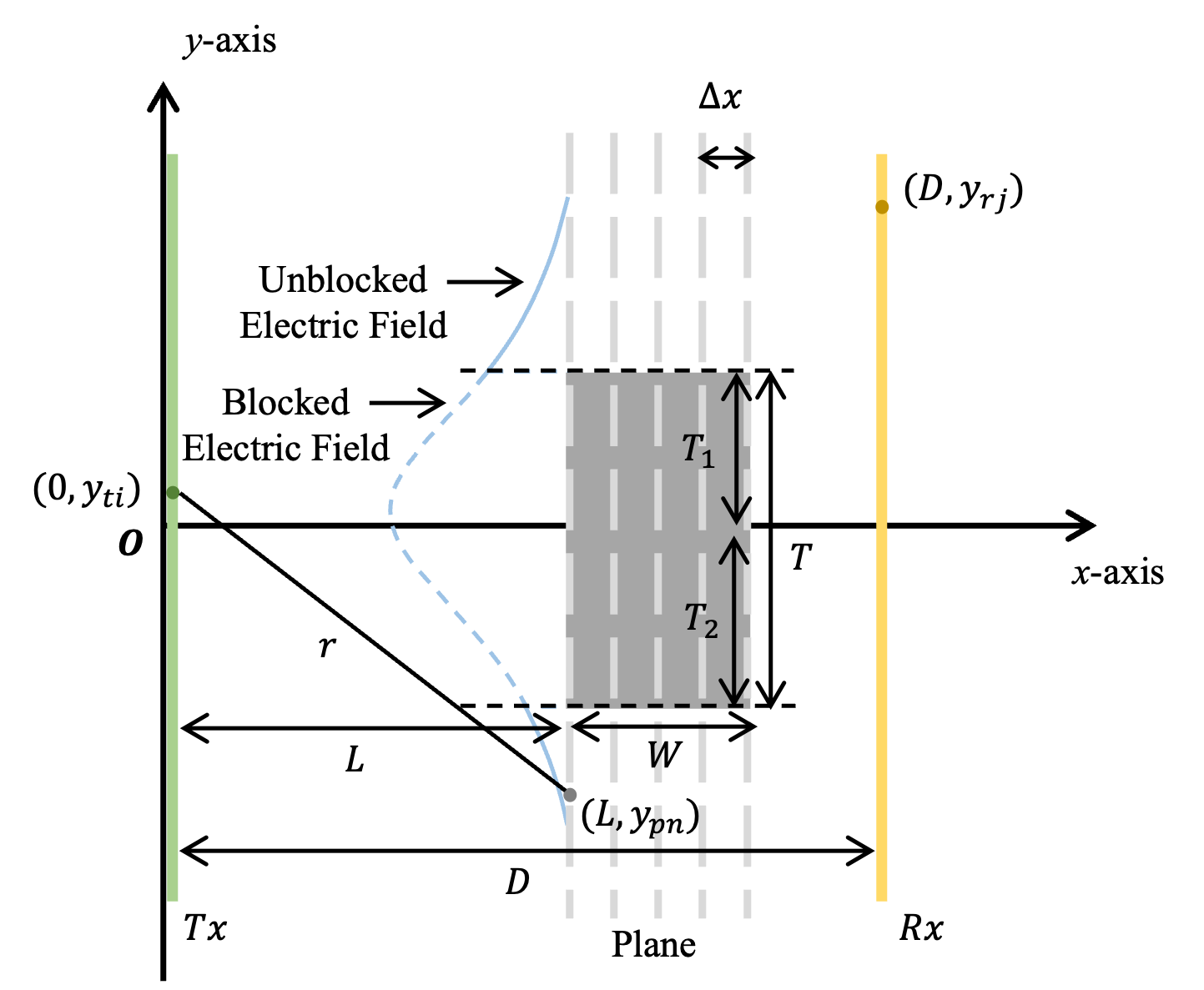}\label{fig:WCM}}
     \subfigure[Cascaded geometric and wave channel model. ]{\includegraphics[width=0.33\linewidth]{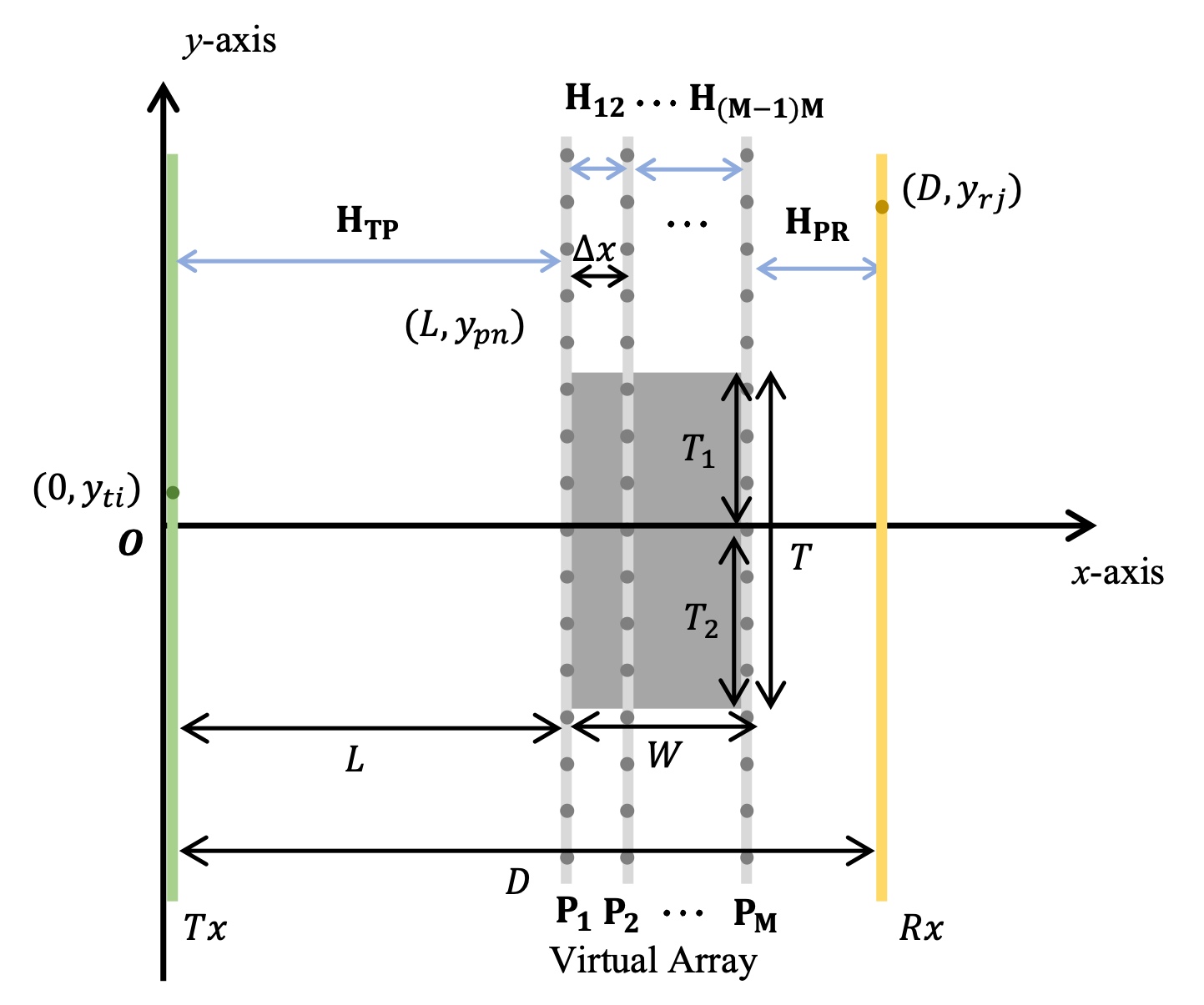}\label{fig:CGWCM}}
   \caption{The accuracy of the GCM and CGWCM.}
\end{figure*}

The GCM models electromagnetic wave propagation as straight-line transmission, neglecting wave effects, and is widely adopted in the literature \cite{Yuan-2023-Spatial}. As shown in Fig. \ref{fig:GCM}, the ULAs are deployed at the Tx and Rx, where $y_{ti}$ and $y_{rj}$ denote the vertical coordinates of the $i^{\mathrm{th}}$ transmitting and $j^{\mathrm{th}}$ receiving antennas, respectively, with $i=1,\cdots,N_t$ and $j=1,\cdots,N_r$. To simplify analysis, the blockage is modeled as a rectangle of width $W$ and length $T = T_1 + T_2$. $D$ denotes the distance between the Tx and Rx and $L$ is the distance from the Tx to the blockage. The four lines extending from the $i^{\mathrm{th}}$ transmitting antenna to the vertices of the blockage are labeled as $l_1$, $l_2$, $l_3$, and $l_4$, intersecting the Rx array plane at points $b_1$, $b_2$, $b_3$, and $b_4$, which can be expressed as
\begin{subequations}
    \begin{align}
        &b_1=\frac{T_1-y_{ti}}{L}D+y_{ti}, \\
        &b_2=\frac{T_1-y_{ti}}{L+W}D+y_{ti}, \\
        &b_3=\frac{-T_2-y_{ti}}{L}D+y_{ti},\\
        &b_4=\frac{-T_2-y_{ti}}{L+W}D+y_{ti}.
    \end{align}
    \label{eq:y_ti}
\end{subequations}
In GCM, the complex path gain from the $i^{\mathrm{th}}$ transmitting antenna to the $j^{\mathrm{th}}$ receiving antenna is denoted as $\alpha^{ij}_g$ and the complex gain of all antenna pairs between the Tx and Rx is computed individually. Since GCM models electromagnetic wave propagation as a straight line, antenna pairs with an unblocked path have a non-zero complex gain, while those with a blocked path have zero gain, meaning the transmitted signal is completely obstructed. In Fig. \ref{fig:GCM}, for a fixed $y_{ti}$, the 
propagation path remains unblocked if $y_{rj} > \text{max}(b_1,b_2)$ or $y_{rj} < \text{min}(b_3,b_4)$. Otherwise, the propagation path is blocked. Consequently, the quasi-LoS channel response between the $i^{\mathrm{th}}$ transmitting antenna and the $j^{\mathrm{th}}$ receiving antenna can be expressed as
\begin{equation}
\label{eq:GCM}
\mathbf{H}_{\mathrm{G}}(i,j) = 
\begin{cases} 
\left|\alpha_{g}^{ij}\right| e^{-j\frac{2\pi}{\lambda}D^{ij}}, & 
    \begin{aligned}
    &y_{rj} > \max(b_{1}, b_{2}) \\
    &\quad \text{or } y_{rj} < \min(b_{3}, b_{4})
    \end{aligned} \\
0, & 
    \begin{aligned}
    &\min(b_{3}, b_{4}) \leq y_{rj} \\
    &\quad \leq \max(b_{1}, b_{2})
    \end{aligned}
\end{cases}
\end{equation}
where $D^{ij}$ represents the communication distance from the $i^{\mathrm{th}}$ transmitting antenna to the $j^{\mathrm{th}}$ receiving antenna and $\mathbf{H}_{\mathrm{G}}$ denotes the $N_r\times N_t$-dimensional geometrical channel matrix. Considering the free spreading loss, the path gain $|\alpha^{ij}_g|$ in \eqref{eq:GCM} can be explicitly written as
 $|\alpha^{ij}_g|=\frac{c}{4\pi f D^{ij}}$, where $c$ is the speed of light and $f$ is the carrier frequency. 
 
 In exchange for its low computational complexity of $\mathcal{O}(N_tN_r)$, GCM has low accuracy in quasi-LoS conditions as it neglects electromagnetic wave diffraction.

% \begin{figure}[t]
%     \centering
%     \includegraphics[width=0.7\linewidth]{images/channel_model/GCM.png} % 使用 \linewidth 自适应栏宽
%     \caption{Geometrical channel model for ULA communication systems.}
%      \label{fig:GCM}
% \end{figure}
 \subsection{WCM}
% \begin{figure}[t]
%     \centering
%     \includegraphics[width=0.7\linewidth]{images/channel_model/WCM.png} % 使用 \linewidth 自适应栏宽
%     \caption{Wave-based channel model for ULA communication systems.}
%      \label{fig:WCM}
% \end{figure}
The WCM is derived from the Huygens-Fresnel principle, which accounts for Electromagnetic (EM) wave diffraction and effectively characterizes wave behavior, particularly in the quasi-LoS \cite{Chen-2024-Curving}. According to this principle, each point on the wavefront acts as a secondary source emitting spherical waves, and the superposition of these secondary waves determines the resulting field at the receiver. To describe wavefront propagation on a plane, the Rayleigh–Sommerfeld (RS) integral can be employed \cite{Chen-2024-Curving}, enabling the calculation of the electric field at any point based on a known input field. Let the $x_0$-plane be defined at $x = x_0$, and $\mathbf{E}(x_0, y_0)$ denote the electric field at position $(x_0, y_0)$. The electric field on the $x_0$-plane is represented as a column vector $\mathbf{E}(x_0, \colon)$, which contains values at all positions $(x_0, y)$ on the plane. Based on the input field $\mathbf{E}(x_{\mathrm{in}}, \colon)$, the electric field at an arbitrary location $(x_0, y_0)$ can be expressed as
\begin{subequations}
    \begin{align}
        \mathbf{E}(x_0,y_0)=&RS(\mathbf{E}(x_\mathrm{in},:),x_0,y_0)\\=&Q_\mathrm{w}\int\mathbf{E}(x_\mathrm{in},y^\prime)\frac{x}{2\pi r^2}e^{ikr}(\frac{1}{r}-ik)dy^\prime,
    \end{align}
\end{subequations}
where $Q_\mathrm{w}$ is a scalar parameter, $x_\mathrm{in}$ is the propagation coordinate of input plane, $y^\prime$ is the transverse coordinate of input electric field, $r=\sqrt{(x_0-x_{in})^2+(y_0-y^\prime)^2}$ and $k=\frac{2\pi}{\lambda}$. 
% For computational simplicity, we can quantify and select $N$ points on the plane, with a spacing of half wavelength between each point and denote the transverse coordinate of $n^{\mathrm{th}}$ point as $y_{pn}$ with $n=1,\cdots,N$. Therefore, the electric field at $(x,y)$ can be derived as
% \begin{subequations}
%     \begin{align}
%         \mathbf{E}(x,y)\approx&RS_q(\mathbf{E}(x_{in},:),x,y)\\
%         =&Q_w\sum_{n=1}^{N}\mathbf{E}(x_{in},y_{pn})\frac{x}{2\pi r_n^2}e^{ikr_n}(\frac{1}{r_n}-ik),
%     \end{align}
% \end{subequations}
% where $r_n=\sqrt{(x-x_{in})^2+(y-y_{pn})^2}$.

As shown in Fig.~\ref{fig:WCM}, the $i^{\mathrm{th}}$ transmitting antenna emits a point electric field $\mathbf{E}(0,y_{ti})$ and forms an electric field column vector $\mathbf{E}(L,\colon)$ on the L-plane, which can be expressed as 
\begin{subequations}
\begin{align}
       \mathbf{E}(L,y_{pn})=&RS(\mathbf{E}(0,y_{ti}),L,y_{pn})\\
       =&Q_w\mathbf{E}(0,y_{ti})\frac{L}{2\pi r_n^2}e^{ikr_n}(\frac{1}{r_n}-ik),
\end{align}
\end{subequations}
where $y_{pn}$ is denoted as the transverse coordinate of $n^{\mathrm{th}}$ point on the plane and $r_n=\sqrt{L^2+(y_{ti}-y_{pn})^2}$. However, because of the presence of blockage, a part of the electric field is blocked and cannot transmit continuously. In \cite{Chen-2024-Curving}, a binary matrix $\mathbf{B}(x,y)$ is introduced, which takes the value of 0 if the blocker includes point $(x,y)$ and takes 1 otherwise. Therefore, the blocked electric field $\mathbf{E}_\mathrm{B}$ on the L-plane can be written as 
\begin{equation}
    \mathbf{E}_\mathrm{B}(L,y_{pn})=\mathbf{B}(L,y_{pn})\mathbf{E}(L,y_{pn}).
\end{equation}
We iteratively apply the RS integral to the planes within the blockage region, with spacing $\Delta x$. After obtaining the electric field $\mathbf{E}_\mathrm{B}(L+W,\colon)$ of $(L+W)$-plane, we can obtain the quasi-LoS channel response between the $i^{\mathrm{th}}$ transmitting antenna and the $j^{\mathrm{th}}$ receiving antenna as
\begin{equation}
    \mathbf{H}_{\mathrm{W}}(i,j)=RS(\mathbf{E}_\mathrm{B}(L+W,:),D,y_{rj}).
\end{equation}
Following the above procedure, the complete channel matrix accounting for the diffraction effect can be obtained. 

Although the WCM provides an accurate characterization of EM wave propagation in the presence of blockages, its integral form leads to high computational complexity, particularly when modeling UM-MIMO channels with a large number of elements.
\subsection{CGWCM}
%我们保留了WCM中通过引入平面来model 衍射的思想，同时鉴于每个平面之间是简单的无障碍物传播，我们使用比RS积分复杂度更低，计算更简洁的GCM去计算平面之间的电磁波传播
% \begin{figure}[t]
%     \centering
%     \includegraphics[width=0.7\linewidth]{images/channel_model/CGWCM.png} % 使用 \linewidth 自适应栏宽
%     \caption{Cascaded geometric and wave channel model for ULA communication systems.}
%      \label{fig:CGWCM}
% \end{figure}
Inspired by both the GCM and WCM, we propose the CGWCM with a simplified form and high accuracy. As shown in Fig.~\ref{fig:CGWCM}, we place $M$ virtual arrays at regular intervals of $\Delta x$ within the blockage region to account for diffraction effects based on the Huygens-Fresnel principle, as in the WCM. The $m^{\mathrm{th}}$ virtual array is denoted as $P_m$, and the transverse coordinate of $n^{\mathrm{th}}$ antenna in the virtual array is denoted as $y_{pn}$.
Each virtual array consists of $N$ antennas with a spacing of $d=\lambda/2$, and the channel between adjacent arrays is modeled using the GCM.
The channel between the Tx and $P_1$ is represented by $\mathbf{H}_{\mathrm{TP}}\in\mathbb{C}^{N\times N_t}$, the channel between $P_m$ and $P_{m+1}$ is denoted by $\mathbf{H}_{m(m+1)}\in\mathbb{C}^{N\times N}$ and the channel between $P_M$ and the Rx is $\mathbf{H}_{\mathrm{PR}}\in\mathbb{C}^{N_r\times N}$. Given the transmitting signal $\mathbf{x}\in\mathbb{C}^{N_t\times1}$, the receiving signal at $P_1$ is $\mathbf{y}_1=\mathbf{H}_{\mathrm{TP}}\mathbf{x}$. Because of the blockage, the antennas within the blocked region cannot transmit the signal.
 Let $\mathbf{B}_\mathrm{m}$ be the binary matrix for the $m^{\mathrm{th}}$ virtual array $P_m$, where entries from row indices $b_{m1}$ to $b_{m2}$ are set to zero, i.e., $\mathbf{B}_\mathrm{m}(b_{m1}:b_{m2},\colon)$ = 0, and the remaining entries are one. Accordingly, the blocked received signal at $P_1$ is given by
\begin{equation}
    \mathbf{y}_{1b}=(\mathbf{B_1}\odot\mathbf{H}_{\mathrm{TP}})\mathbf{x},
\end{equation}
where the dimensional of $\mathbf{B}_1$ is the same as $\mathbf{H}_{\mathrm{TP}}$, i.e., $\mathbf{B}_1\in\mathbb{C}^{N\times N_t}$. Then, $P_1$ transmits the signal to the next virtual array and repeats the similar processing.  After cascading through the virtual arrays, the final received signal $\mathbf{y}_b$ can be expressed as
\begin{equation}
    \mathbf{y}_b=Q_c\mathbf{H}_{\mathrm{PR}}(\mathbf{B}_{\mathrm{M}}\odot\mathbf{H}_{\mathrm{(M-1)M}})\cdots(\mathbf{B}_2\odot\mathbf{H}_{12})(\mathbf{B}_1\odot\mathbf{H}_{\mathrm{TP}})\mathbf{x},
\end{equation}
where $Q_c$ is a scalar parameter and $\mathbf{B}_2$, $\cdots$, $\mathbf{B}_{\mathrm{M}}$ are $N\times N$ dimensional. Therefore, the quasi-LoS channel matrix under the proposed CGWCM framework can be given by 
\begin{equation}
   \mathbf{H}_{\mathbf{C}}= Q_c\mathbf{H}_{\mathrm{PR}}(\prod_{m=2}^{M}\mathbf{B}_{\mathrm{m}}\odot\mathbf{H}_{\mathrm{(m-1)m}})(\mathbf{B}_1\odot\mathbf{H}_{\mathrm{TP}}).
\end{equation}

Compared with WCM, which computes RS integrals individually for each antenna pair with a complexity of $\mathcal{O}(N_tN_rMN^2)$, CGWCM uses matrix operations and achieves a more efficient and parallelizable formulation with a total complexity of $\mathcal{O}(MN^3+N^2N_t+NN_tN_r)$, making it well-suited for large-scale THz UM-MIMO systems.

% The CGWCM effectively combines the simplicity of the GCM with the diffraction modeling capability of the WCM by introducing virtual arrays and blockage-aware binary matrices. This approach gives a simplified form while maintaining high accuracy, making it suitable for THz UM-MIMO systems with a large number of antennas.
\subsection{Calibration}
In LoS scenarios, the GCM provides highly accurate channel modeling. However, in quasi-LoS scenarios, CGWCM introduces virtual arrays to account for diffraction, which leads to inaccuracies in the complex path gain. To address this, we propose a calibration method that corrects the path gain of CGWCM by leveraging the accurate gain provided by the GCM in the LoS case.
Specifically, we introduce virtual arrays at the same locations as those used in the quasi-LoS CGWCM model and compute the LoS channel using CGWCM, denoted as $\mathbf{H}_{\mathrm{C}}^{\mathrm{LoS}}$. We then compute the LoS channel using the GCM, denoted as $\mathbf{H}_{\mathrm{G}}^{\mathrm{LoS}}$, which serves as a reference for accurate complex path gain. By comparing these two channels, we derive the amplitude and phase calibration parameters as 
\begin{subequations}
    \begin{align}
 \alpha_{cal}&=\frac{||\mathbf{H}_{\mathrm{G}}^{\mathrm{LoS}}||_F}{||\mathbf{H}_{\mathrm{C}}^{\mathrm{LoS}}||_F},\\
    \phi_{cal}&=\frac{1}{N_rN_t}\sum_{i=1}^{N_r}\sum_{j=1}^{N_t}\angle\frac{\mathbf{H}_{\mathrm{G}}^{\mathrm{LoS}}(i,j)}{\mathbf{H}_{\mathrm{C}}^{\mathrm{LoS}}(i,j)}. 
\end{align}
\end{subequations}
Then the calibrated LoS and quasi-LoS channel can be given by $\mathbf{H}_{\mathrm{Gc}}^{LoS}=\alpha_{cal}e^{j \phi_{cal}}\mathbf{H}_{\mathrm{G}}^{\mathrm{LoS}}$ and $\mathbf{H}_{\mathrm{Gc}}^{qLoS}=\alpha_{cal}e^{j \phi_{cal}}\mathbf{H}_{\mathrm{G}}^{qLoS}$.
This calibration effectively compensates for the gain mismatch introduced by virtual arrays and ensures we can practically evaluate the spectral efficiency in the quasi-LoS scenarios.

\section{Analysis of Airy Beam Characteristics and beamforming design}
\label{Sec:Analysis of Airy}
\begin{figure}[t]
    \centering
    \subfigure[Free-space propagation.]{\includegraphics[width=\linewidth]{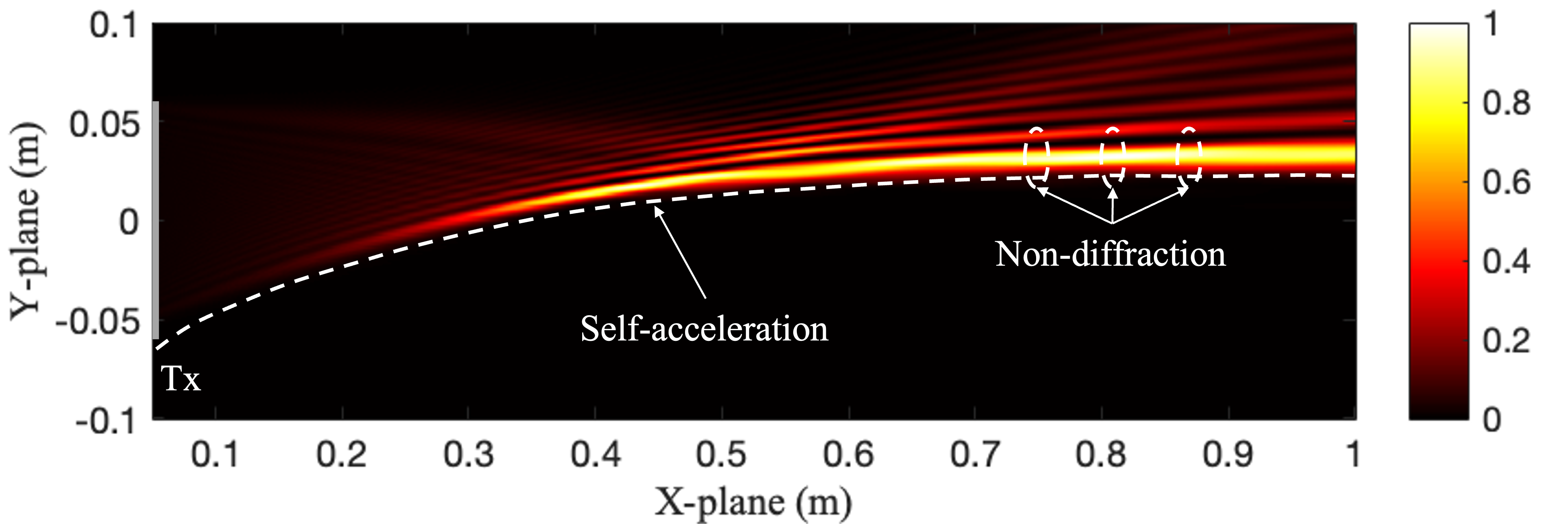}\label{fig:Airy_free_space}} 
     \subfigure[Obstructed propagation.]{\includegraphics[width=\linewidth]{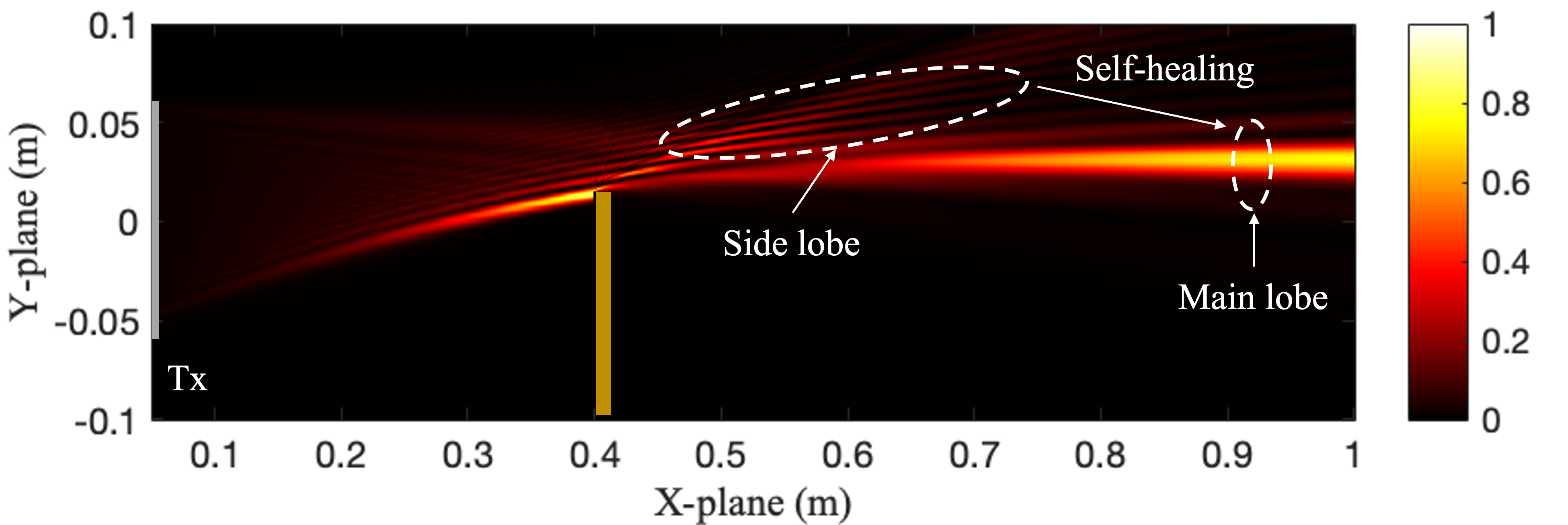}\label{fig:Airy_obstructed}}
    \caption{Propagation characteristics of Airy beam in both free-space and obstructed propagation.}
\end{figure}
In this section, we first investigate the characteristics of the Airy beam in the quasi-LoS. Then, a comprehensive discussion is provided on its beamforming design.
\subsection{Propagation Characteristics}
% Actually, the Airy beam has been recently studied at optics frequencies, primarily focusing on its characterization \cite{}, which is also applicable to the THz band.
\subsubsection{Non-diffraction}
As illustrated in Fig. \ref{fig:Airy_free_space}, the main lobe of the Airy Beam demonstrates minimal spreading during propagation, which is called a non-diffraction property. This allows the Airy beam to maintain its energy concentration over long distance without significant dispersion, unlike a focusing beam, which concentrates energy at a certain point and fast diffracts resulting in a low tolerance for misalignment in practical communication scenarios. The Airy beam distributes energy along an extended trajectory, which reduces the dependency on precise receiver positioning and enhances the fault tolerance and robustness of the communication system in complex environments.
\subsubsection{Self-acceleration}
As shown in Fig. \ref{fig:Airy_free_space}, the main lobe of the Airy beam does not propagate along a straight line but instead follows a curved trajectory, i.e., self-acceleration. This unique bending property enables the Airy beam to autonomously change its propagation direction during transmission. As a result, it can effectively circumvent blockages in the communication path without significant loss of energy. This capability makes it highly suitable for applications in dynamic complex communication systems.
\subsubsection{Self-healing}
As shown in Fig. \ref{fig:Airy_obstructed}, when the main lobe of the Airy beam is obstructed, the beam gradually reconstructs its original shape after a certain propagation distance, maintaining its self-accelerating trajectory with minimal energy loss. This phenomenon, known as self-healing, is primarily enabled by the side lobes, which carry residual energy to reconstruct the main lobe after being obstructed. This property enhances the robustness of communication quality in complex obstructed scenarios.

%从jornet的curve文章中找引用
\subsection{Airy Beam-based Beamforming Design}
\begin{figure}[t]
    \centering
    \includegraphics[width=0.8\linewidth]{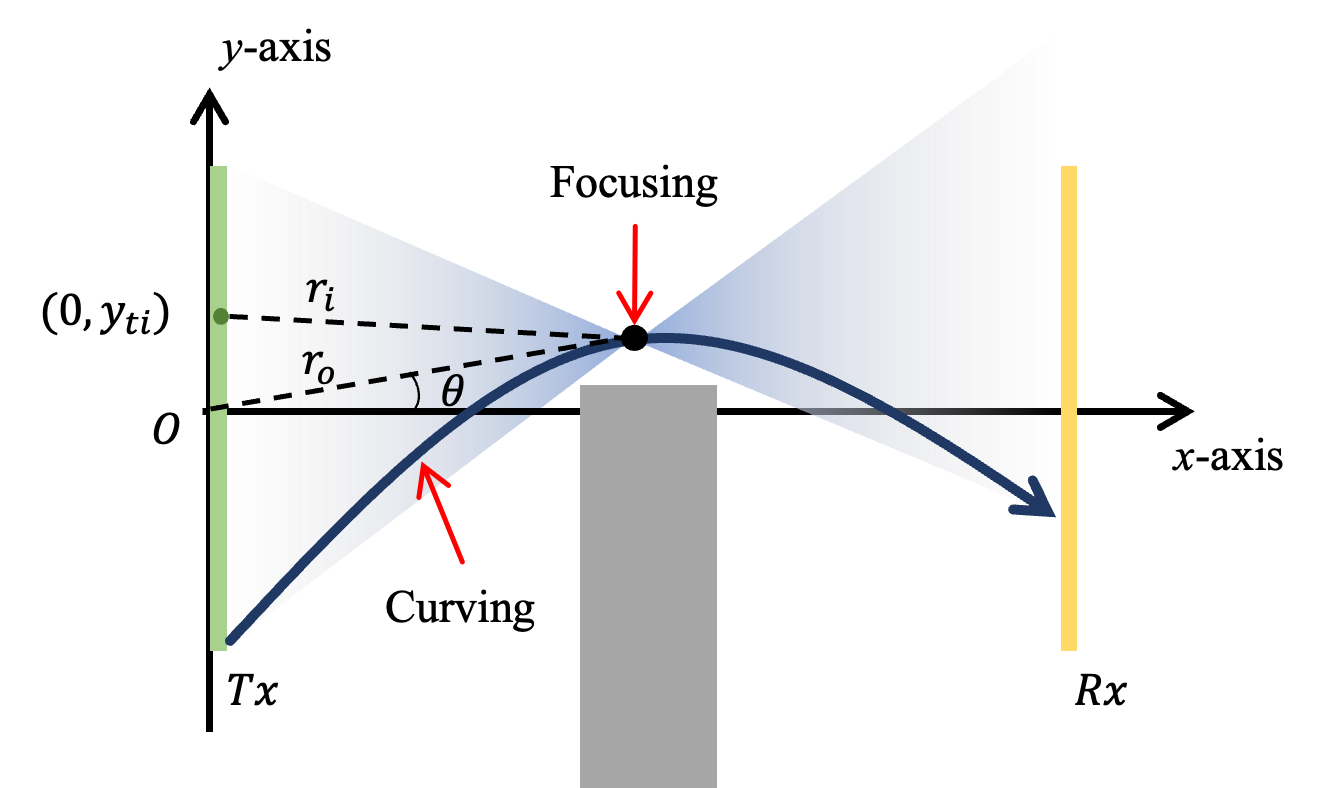} % 使用 \linewidth 自适应栏宽
    \caption{Airy beam generation for ULA communication systems.}
     \label{fig:beam_gen}
\end{figure}
\begin{figure}[t]
    \centering
    \subfigure[$a=0$]{\includegraphics[width=0.37\linewidth]{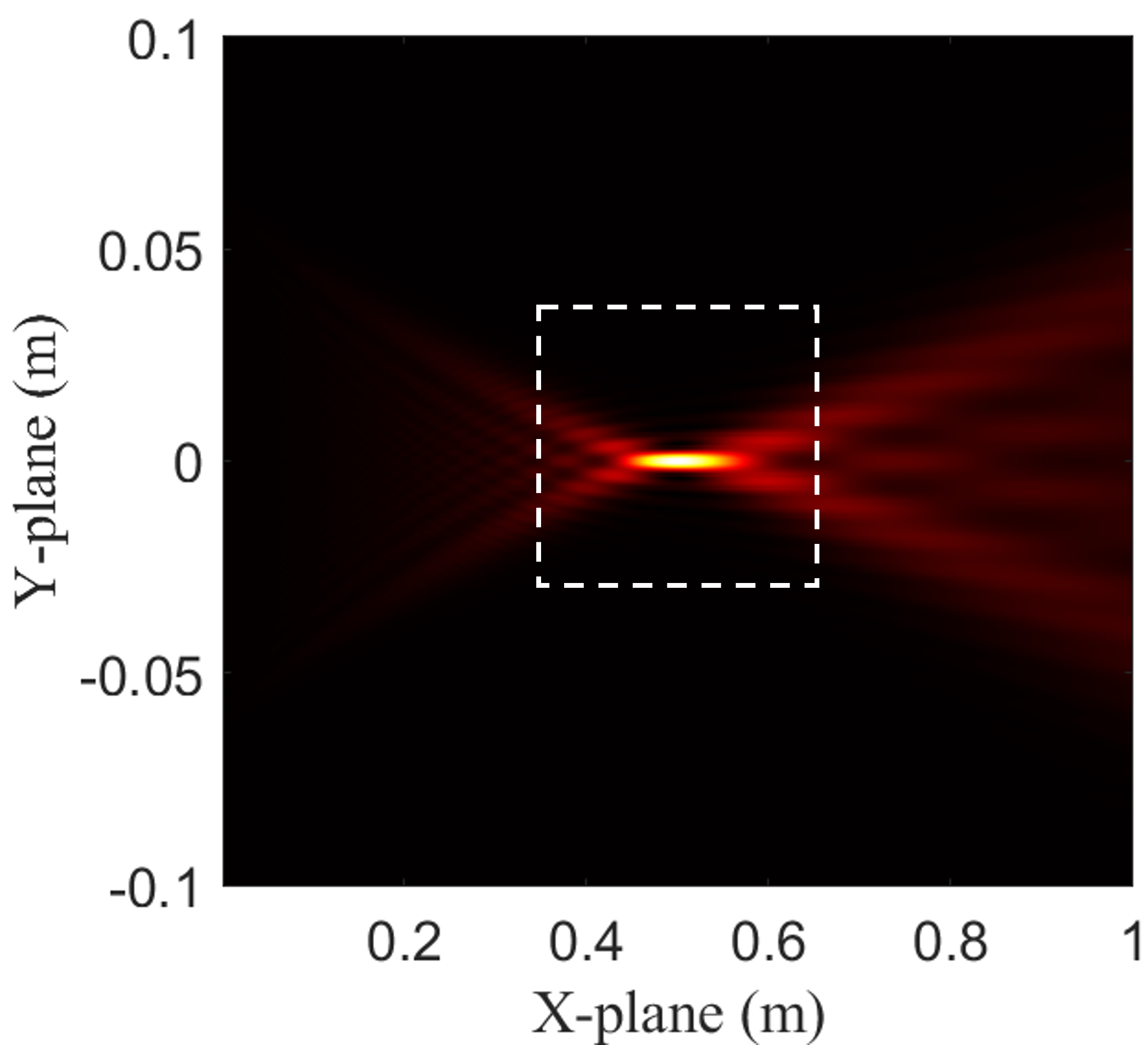}\label{fig:Airy_a0}} 
     \subfigure[$a=2$]{\includegraphics[width=0.37\linewidth]{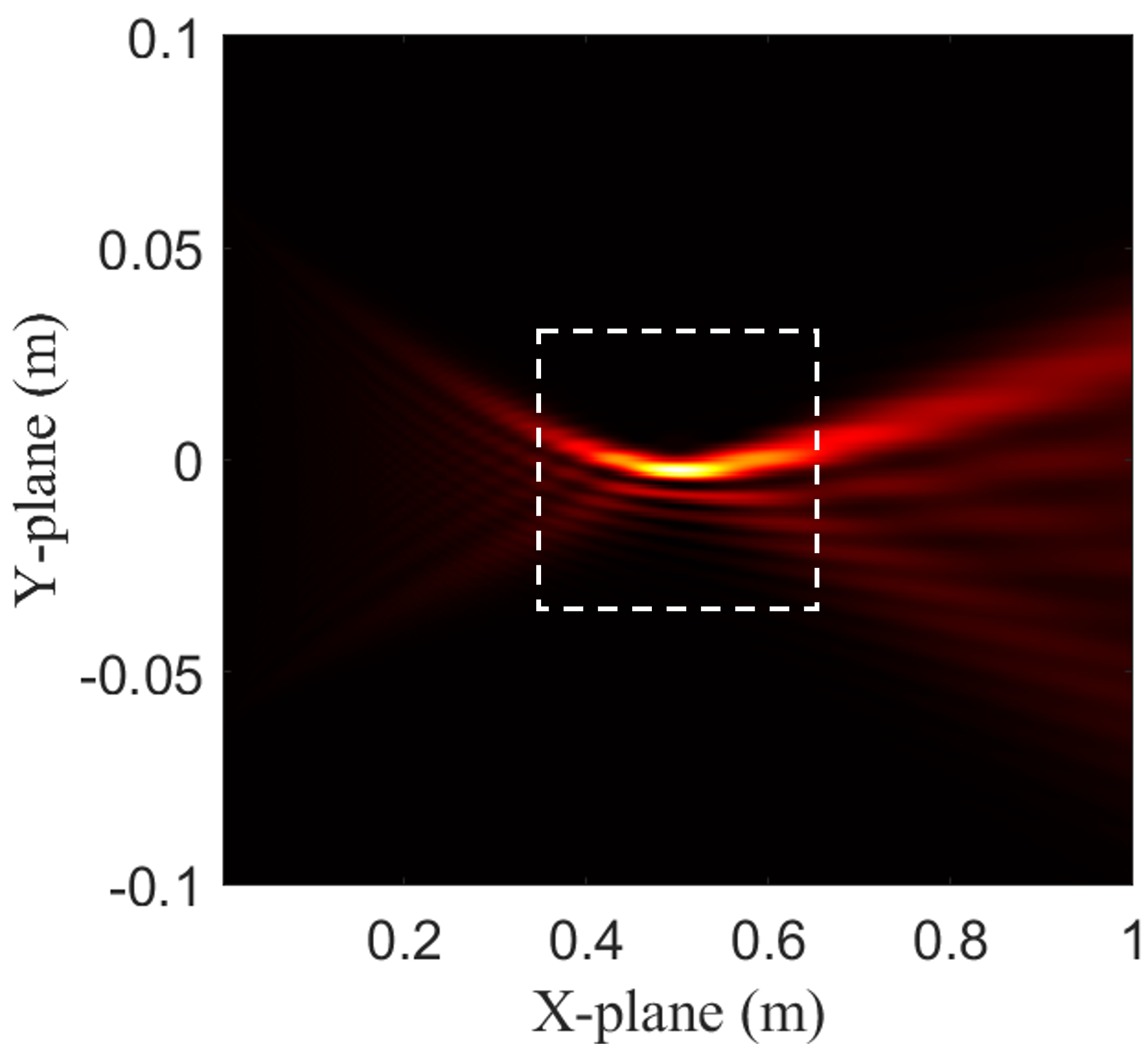}\label{fig:Airy_a2}}
     \\
     \subfigure[$a=-2$]{\includegraphics[width=0.37\linewidth]{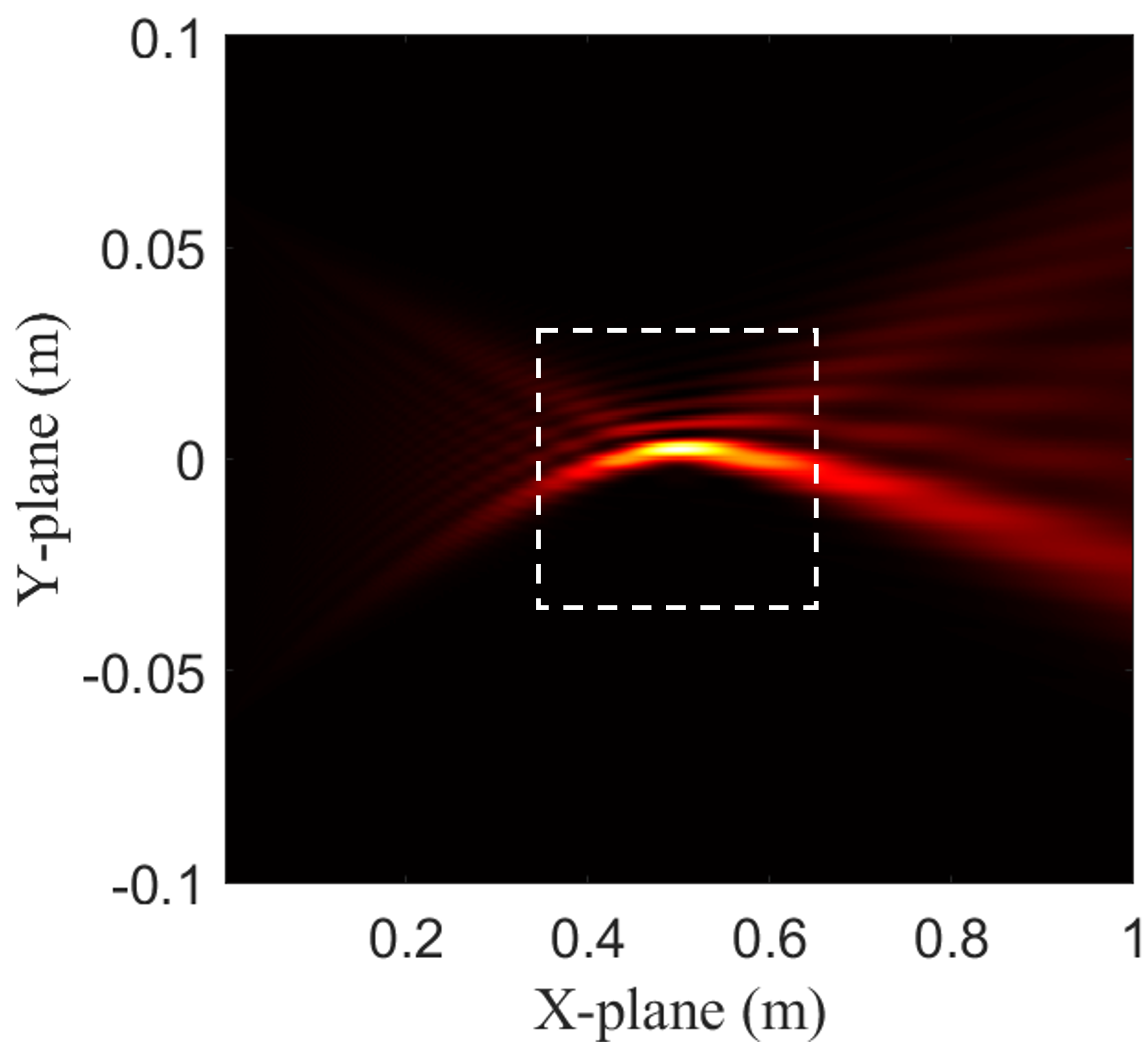}\label{fig:Airy_a-2}}
     \subfigure[$a=-4$]{\includegraphics[width=0.37\linewidth]{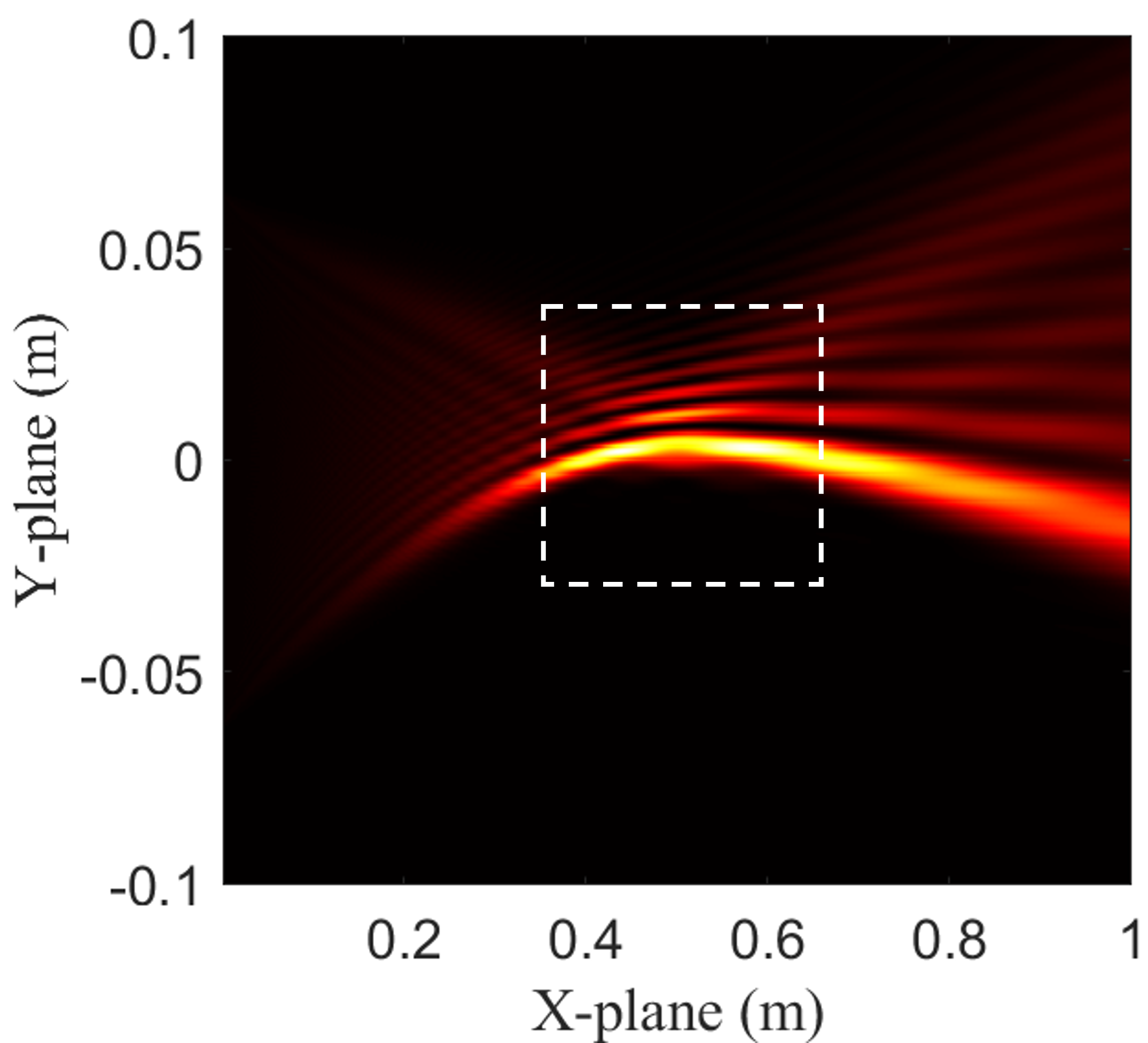}\label{fig:Airy_a-4}}
   \caption{Airy beams focusing on the point $(0.5,0)$ with different values of the cubic coefficient $a$.}
   \label{fig:Airy_focus_beam}
\end{figure}
%%图片中加个箭头表示图片之间的比较
\subsubsection{Airy beam analog precoder}There are two main approaches to generating an Airy beam. The first method involves directly modulating the complex amplitude in each antenna, which can be expressed as 
\begin{equation}
E(y_{ti})=Ai(y_{ti}/y_o)\exp({by_{ti}/y_o}),
\end{equation}
where $Ai(\cdot)$ denotes the Airy function, $b$ is a truncation factor that ensures energy containment, and $y_o$ controls the degree of curvature of the trajectory \cite{Petrov-2024-Wavefront-Hopping}. While this approach allows for the creation of an arbitrary Airy beam, it is impractical for THz UM-MIMO systems due to the difficulty in controlling both the amplitude and phase of each antenna. However, the Fourier transform of the Airy wave solution is given by a Gaussian beam with an additional cubic phase, which means that we can generate an Airy beam by simply imposing a cubic phase to a Gaussian beam using lens or phase shifters. 

The second method, which is more suitable for THz hybrid beamforming communication systems, involves adjusting the phase profile of the Airy beam, given by
\begin{equation}
    \phi_{a}(i)=\frac{2\pi}{\lambda}(ay_{ti}^3+by_{ti}^2+cy_{ti}),
\end{equation}
where $a$, $b$, and $c$ are cubic, quadratic, and linear coefficients, respectively. In the near-field, the Gaussian beam exhibits focusing property, which is determined by quadratic and linear coefficients \cite{Zhang-2022-Beam-focusing}. Therefore, the Airy beams generated by adding cubic phase to Gaussian beams exhibit both curved and focusing properties. As shown in Fig. \ref{fig:beam_gen}, the near-field focusing phase for $i^{\mathrm{th}}$ antenna of ULA can be expressed as 
\begin{equation}
    \phi_{f}(i)=\frac{2\pi}{\lambda}(r_i-r_o),
\end{equation}
%引用DDL的文章LDMA
where $r_i$ denotes the distance between the focusing point and the $i^{\mathrm{th}}$ antenna, and $r_o$ is the distance between the focusing point and the center of the Tx array. The distance term $r_i$ can be written based on the cosine theorem as
\begin{subequations}
\begin{align}
    r_i=&\sqrt{r_o^2+y_{ti}^2-2r_oy_{ti}\sin{\theta}}\\
    =&r_o\sqrt{1+(\frac{y_{ti}^2}{r_o^2}-\frac{2y_{ti}\sin{\theta}}{r_o})}\\
    \approx&r_o-\sin{\theta}y_{ti}+\frac{\cos^2{\theta}}{2r_o}y_{ti}^2.\label{eq:focus_approx}
\end{align}
\end{subequations}
The approximation in \eqref{eq:focus_approx} is derived by Talor series expansion$\sqrt{1+z}=1+\frac{z}{2}-\frac{z^2}{8}+\mathcal{O}(z^2)$ where $z=\frac{y_{ti}^2}{r_o^2}-\frac{2y_{ti}\sin{\theta}}{r_o}$ and retaining the first and second-order of $\frac{y_{ti}}{r_o}$, as detailed in \cite{Wu-2023-Multiple}. Then, $\phi_{f}$ can be rewritten as
\begin{equation}
    \phi_{f}(i)=\frac{2\pi}{\lambda}(\frac{\cos^2{\theta}}{2r_o}y_{ti}^2-\sin{\theta}y_{ti}).
\end{equation}
Next, we add the cubic phase $ay_t^3$ to the phase profile of the Gaussian beam $\phi_{f}$ so that the phase profile of the Airy beam can be given by
\begin{equation}
    \phi_a(i)=\frac{2\pi}{\lambda}(ay_{ti}^3+\frac{\cos^2{\theta}}{2r_o}y_{ti}^2-\sin{\theta}y_{ti}),
\end{equation}
where $a$ introduces the self-acceleration property. Therefore, we can generate an arbitrary Airy beam by adjusting the cubic phase coefficient and the focusing point. The Airy beams with different values of the cubic coefficient $a$ are presented in Fig.~\ref{fig:Airy_focus_beam}. Compared Fig.~\ref{fig:Airy_a0} with Fig.~\ref{fig:Airy_a2}, \ref{fig:Airy_a-2}, and \ref{fig:Airy_a-4}, we observe that the energy of the Airy beams consistently concentrates around the focusing point $(0.5,0)$, regardless of the cubic coefficient $a$. A closer comparison of Fig.~\ref{fig:Airy_a2}, \ref{fig:Airy_a-2}, and \ref{fig:Airy_a-4} shows that the cubic coefficient $a$ plays a critical role in shaping the propagation characteristics of the Airy beams. The sign of $a$ determines the direction of bending. When $a>0$, the beam follows a concave trajectory. When $a<0$, the beam follows a convex trajectory. The absolute value of $a$ influences both the curvature and the energy distribution of the beam. A larger $|a|$ results in a more pronounced curvature of the beam and a greater spatial extension of the energy. These observations provide guidelines for tailoring Airy beams with desired propagation properties, which can be leveraged in complex applications and environments.

\subsubsection{Digital precoder and combiner }After we obain the Airy beam analog precoder $\mathbf{F}_{\mathrm{A}}$. Given the channel matrix $\mathbf{H}$, we can obtain the effective channel between the Tx RF chains and Rx antennas, i.e., $\mathbf{H}_{\mathrm{eff}}=\mathbf{H}\mathbf{F}_{\mathrm{A}}$. Then the precoder and combiner can be derived based on Singular Value Decomposition (SVD) as 
\begin{subequations}
\label{eq:precoder}
    \begin{align}
        \mathbf{H}_\mathrm{eff}=&\mathbf{U}\Sigma\mathbf{V}^\mathrm{H},\\
        \mathbf{F}_\mathrm{BB}=&\mathbf{V}(:,1:L_t),\\
        \mathbf{W}_\mathrm{opt}=&\mathbf{U}(:,1:L_r),
    \end{align}
\end{subequations}
where $\mathbf{F}_\mathrm{BB}$ and $\mathbf{W}_\mathrm{opt}$ are the first $L_t$ and $L_r$ columns of $\mathbf{U}$ and $\mathbf{V}$, respectively. Given the optimal combiner $\mathbf{W}_\mathrm{opt}$, the analog combiner $\mathbf{W}_\mathrm{RF}$ and digital combiner $\mathbf{W}_\mathrm{BB}$ can be obtained by solving a Euclidean distance minimization problem, i.e., $||\mathbf{W}_\mathrm{opt}-\mathbf{W}_\mathrm{RF}\mathbf{W}_\mathrm{BB}||_\mathrm{F}$ which is well-studied in the literature \cite{Ayach-2014-Spatially}.

\section{Airy Beam Codebook Design in Quasi-LoS Scenarios}
%加回来。。。。
\label{sec:beam search}
In this section, we first analyze the correlation between two arbitrary Airy beams in a ULA-based system. Based on this, we propose beam search schemes and codebook designs with complexity-performance trade-offs, enabling efficient Airy beam selection without prior knowledge of blockage position.
% We begin with the exhaustive search codebook, which provides optimal performance with Airy beam at the cost of high searching overhead and computational complexity. To address this limitation, we introduce the focusing-Airy codebook, a favorable balance between performance and implementation complexity. Finally. we propose a low-complexity codebook design that significantly reduce the overhead while maintaining a satisfactory performance. 

\subsection{Correlation Analysis of Airy Beam Vectors}
% \begin{figure}[t]
%     \centering
% {\includegraphics[width=\linewidth]{images/correlation/Airy_focus.png}\label{fig:correlation}}           \caption{Airy beams focusing on the point $(0.5,0)$ with different values of the cubic coefficient $a$.}
%    \label{fig:Airy_focus_beam}
% \end{figure}
As demonstrated in Sec. \ref{Sec:Analysis of Airy}, the phase vector of an arbitrary Airy beam in a ULA system is characterized by three fundamental parameters: the cubic coefficient $a$, the distance $r$ and the angle $\theta$. These parameters collectively determine the Airy beam and can be expressed as
\begin{equation}
    \mathbf{a}(a,r,\theta)=\frac{1}{\sqrt{N_t}}[\Phi^1,\cdots,\Phi^{N_t}]^T,
\end{equation}
\begin{equation}
   \Phi^i= e^{j\frac{2\pi}{\lambda}(ay_{ti}^3+\frac{\cos^2{\theta}}{2r}y_{ti}^2-\sin{\theta}y_{ti})}.
\end{equation}
Then, the correlation of two arbitrary Airy beams with distinct cubic coefficients $a_m$ and $a_n$, focusing on polar coordinates $(r_m,\theta_m)$ and $(r_n,\theta_n)$, respectively, can be formulated as 
\begin{subequations}
\begin{align}
     C=&|\mathbf{a}^H(r_m,\theta_m,a_m)\mathbf{a}(r_n,\theta_n,a_n)|\\
     =&\frac{1}{N_t}\Bigg|\sum_{i=1}^{N_t}\text{exp}\Bigg(jk\Big((a_m-a_n)y_{ti}^3 \nonumber\\&+(\frac{\cos^2{\theta_m}}{2r_m}-\frac{\cos^2{\theta_n}}{2r_n})y_{ti}^2+(\sin{\theta_n}-\sin{\theta_m})y_{ti}\Big)\Bigg)\Bigg|,\label{eq:y_ti_cor}
\end{align}
\end{subequations}
where $y_{ti}$ is the vertical coordinate of $i^{\mathrm{th}}$ transmitting antenna defined as in \eqref{eq:y_ti}.

We consider a symmetric antenna array configuration about the origin with an inter-element spacing of $d$. The antenna positions can be expressed as $y_{ti}=(i-\frac{N_t+1}{2})d$ for $i=1,2,\cdots,N_t$ and the array spans from $-\frac{N_t-1}{2}d$ to $\frac{N_t-1}{2}d$. Then, the correlation expression can be rewritten as
\begin{align}
    C =&\frac{1}{N_t}\Bigg|\sum_{n=-\frac{M}{2}}^{\frac{M}{2}}\text{exp}\Bigg(jk\Big((a_m-a_n)(nd)^3 \nonumber\\  &+(\frac{\cos^2{\theta_m}}{2r_m}-\frac{\cos^2{\theta_n}}{2r_n})(nd)^2
    +(\sin{\theta_n}-\sin{\theta_m})nd\Big)\Bigg)\Bigg|. 
\end{align}
The discrete summation can be approximated as a continuous integral when the number of antennas $N_t$ is sufficiently large. This approximation, based on the Riemann sum concept, becomes increasingly accurate as the inter-element spacing $d$ decreases relative to the array size, which can be expressed as
    \begin{align}
    \label{eq:integral}
           C&\approx \frac{1}{N_t} \Bigg|\int_{-\frac{N_t}{2}}^{\frac{N_t}{2}}\text{exp}\Bigg( jk\Big((a_m-a_n)d^3n^3\nonumber\\&+(\frac{\cos^2{\theta_m}}{2r_m}-\frac{\cos^2{\theta_n}}{2r_n})d^2n^2
    +(\sin{\theta_n}-\sin{\theta_m})dn\Big)\Bigg)\text{d}n\Bigg|.
    \end{align}
It appears challenging to directly derive the sampling scheme for the cubic coefficient of the curving characteristics as well as the distance and angle parameters of focusing characteristics from \eqref{eq:integral}. To simplify the analysis, we can focus on each part individually and sequentially.
\subsubsection{Curving coefficient}
To derive the sample interval for the curving coefficient $a$, we can analyze the cubic component $k(a_m-a_n)d^3$. When considering two arbitrary Airy beams that focus on the same point, i.e., $r_m=r_n$ and $\theta_m=\theta_n$, both the quadratic component $(\frac{\cos^2{\theta_m}}{2r_m}-\frac{\cos^2{\theta_n}}{2r_n})d^2$ and the linear component $(\sin{\theta_n}-\sin{\theta_m})d$ in \eqref{eq:integral} become 0. Consequently, the beam correlation becomes exclusively dependent on the cubic component, which can be expressed as
\begin{subequations}
    \begin{align}
          C_{a_m,a_n}&\approx \frac{1}{N_t}\Bigg|\int_{-\frac{N_t}{2}}^{\frac{N_t}{2}}\text{exp}\Big(j\frac{2\pi}{\lambda}(a_m-a_n)d^3n^3\Big)\text{d}n\Bigg|\\
        &= \frac{1}{N_t}\Bigg|\int_{-\frac{N_t}{2}}^{\frac{N_t}{2}}\text{exp}(j\pi\alpha n^3) \text{d}n\Bigg|,
    \end{align}
\end{subequations}
where $\alpha=\frac{2}{\lambda}(a_m-a_n)d^3$. If $\alpha\geq 0$, i.e., $a_m\geq a_n$, the correlation can be rewritten as
\begin{subequations}
    \begin{align}
        C_{a_m,a_n}
        &\approx \frac{1}{N_t}\Bigg|\int_{-\frac{N_t}{2}}^{\frac{N_t}{2}}\text{exp}(j\pi\alpha n^3) \text{d}n\Bigg|\\
        &=\frac{1}{\sqrt[3]{2\alpha}{N}}\Bigg|\int_{-\sqrt[3]{2\alpha}{N}/{2}}^{\sqrt[3]{2\alpha}{N}/{2}}\text{exp}(j\frac{\pi}{2}t^3)\text{d}t\Bigg|\\
        &=\Bigg|\frac{2\int_{0}^{\sqrt[3]{2\alpha}{N}/{2}}\cos(\frac{\pi}{2}t^3)\text{d}t+j\int_{-\sqrt[3]{2\alpha}{N}/{2}}^{\sqrt[3]{2\alpha}{N}/{2}}\sin(\frac{\pi}{2}t^3)\text{d}t}{\sqrt[3]{2\alpha}{N}}\Bigg| \label{eq:oula}\\
        &=\Bigg|\frac{\int_{0}^{\sqrt[3]{2\alpha}{N}/{2}}\cos(\frac{\pi}{2}t^3)\text{d}t}{\sqrt[3]{2\alpha}{N}/2}\Bigg|\label{eq:cubic_result},
    \end{align}\label{eq:cubic_corr}
\end{subequations}
where $t$ is denoted as $t=\sqrt[3]{2\alpha}n$ and \eqref{eq:oula} comes from Euler's formula. Since $\sin{\frac{\pi}{2}t^3}$ and $\cos{\frac{\pi}{2}t^3}$ are even and odd functions, respectively, \eqref{eq:cubic_result} can be obtained by utilizing the properties of odd and even functions in symmetric integrals. If $\alpha<0, i.e., a_m < a_n$, we denote $t=-\sqrt[3]{2\alpha}n$ and the derivation of correlation is similar to \eqref{eq:cubic_corr}, which can be expressed as
\begin{equation}
    C_{a_m,a_n}\approx\Bigg|\frac{\int_{0}^{-\sqrt[3]{2\alpha}{N_t}/{2}}\cos(\frac{\pi}{2}t^3)\text{d}t}{-\sqrt[3]{2\alpha}{N_t}/2}\Bigg|.
\end{equation}
Therefore, the correlation of two Airy beams focusing on the same point can be formulated as
\begin{subequations}
\begin{align}
    C_{a_m,a_n}(\overline{\alpha})&\approx\Bigg|\frac{\int_{0}^{\sqrt[3]{2|\alpha|}{N_t}/{2}}\cos(\frac{\pi}{2}t^3)\text{d}t}{\sqrt[3]{2|\alpha|}{N_t}/2}\Bigg|\\
    &=\Big|\frac{A(\overline{\alpha})}{\overline{\alpha}}\Big|,\label{eq:curving_corr}
\end{align}
\end{subequations}
where $\overline{\alpha}=\sqrt[3]{2|\alpha|}N_t/2$ and $A(x)=\int_0^{x}\cos(\frac{\pi}{2}t^3)dt$. This relationship allows the parameter $\overline{\alpha}$ corresponding to a target beam correlation $C_{a_m,a_n}$ to be determined by solving \eqref{eq:curving_corr}. The resulting $\overline{\alpha}$ then guides the selection of the curving coefficient interval design.
%针对二次项，引用DDL的论文，然后简要分析直接给出表达式
\subsubsection{Distance coefficient}
Secondly, to derive the sample interval for the distance coefficient $r$, we should focus on the quadratic component $(\frac{\cos^2{\theta_m}}{2r_m}-\frac{\cos^2{\theta_n}}{2r_n})d^2n^2$. When considering two arbitrary Airy beams with the same curving coefficient and angle of focusing point, i.e., $a_m=a_n$ and $\theta_m=\theta_n$, both the cubic component $(a_m-a_n)d^3$ and the linear component $(\sin{\theta_n}-\sin{\theta_m})d$ become 0. The beam correlation depending on the quadratic component can be given  by\cite{Wu-2023-Multiple}
\begin{subequations}
    \begin{align}
        C_{r_m,r_n}(\overline{\beta})&\approx \frac{1}{N_t}\Bigg| \int_{-\frac{N_t}{2}}^{\frac{N_t}{2}}\text{exp}\left(j\frac{\pi}{\lambda}(\frac{1}{r_m}-\frac{1}{r_n}) d^2n^2\cos^2\theta \right) \text{d}n\Bigg|\\
        &=\frac{1}{N_t}\Bigg|\int_{-\frac{N_t}{2}}^{\frac{N_t}{2}}\text{exp}(j\pi\beta n^2)\text{d}n\Bigg|\\
       % &=\frac{1}{\sqrt{2\beta}N_t}|\int_{-\sqrt{2\beta}N_t/2}^{\sqrt{2\beta}N_t/2} \text{exp}(j\frac{\pi}{2}t^2)\text{d}t|\\
        &=\Bigg|\frac{\int_{0}^{\sqrt{2\beta}N_t/2}\cos(\frac{\pi}{2}t^2)+j\int_{0}^{\sqrt{2\beta}N_t/2}\sin(\frac{\pi}{2}t^2)\text{d}t}{\sqrt{2\beta}N_t/2}\Bigg|\\
        &=\Bigg|\frac{B(\overline{\beta})+jD(\overline{\beta})}{\overline{\beta}}\Bigg|,
    \end{align}
\end{subequations}
where $t=\sqrt{2\beta}n$, $\beta=\frac{\cos^2\theta}{\lambda}|\frac{1}{r_m}-\frac{1}{r_n}|d^2$, $\overline{\beta}=\sqrt{2\beta}N_t/2$, $B(x)=\int_0^x\cos(\frac{\pi}{2}t^2)\text{d}t$ and $D(x)=\int_0^x\sin(\frac{\pi}{2}t^2)\text{d}t$ which are Fresnel functions\cite{Wu-2023-Multiple}. 
Then, $\overline{\beta}$ can be determined from a target correlation $C_{r_m,r_n}$, enabling distance coefficient design.

\subsubsection{Angle coefficient}
For the angle coefficient $\theta$, we should focus on the linear component $(\sin{\theta_n}-\sin{\theta_m})dn$. We let the cubic and quadratic components equal to 0, i.e., $a_m=0$ and $\frac{\cos^2\theta_m}{2r_m}=\frac{\cos^2\theta_n}{2r_n}$, which means that two Airy beams with the same curving coefficient and the focusing points are on the curve $\frac{\cos^2\theta}{r}=l$\cite{Cui-2022-Channel-estimation}. Then the beam correlation depending on the linear component can be given as
\begin{subequations}
    \begin{align}
        C_{\theta_m,\theta_n}(\overline{\gamma})=&\frac{1}{N_t}\Bigg|\sum_{n=-\frac{N}{2}}^{\frac{N}{2}}\text{exp}(j\frac{2\pi}{\lambda}(\sin{\theta_n}-\sin{\theta_m})nd)\Bigg| \\
        =&\Bigg|\frac{\sin(\frac{N_t\overline{\gamma}}{2})}{N_t\sin(\frac{\overline{\gamma}}{2})}\Bigg|,
    \end{align}
\end{subequations}
% \begin{subequations}
%     \begin{align}
%         C_{\theta_m,\theta_n}=&\frac{1}{N_t}|\sum_{n=-\frac{N}{2}}^{\frac{N}{2}}\text{exp}(j\frac{2\pi}{\lambda}(\sin{\theta_n}-\sin{\theta_m})nd)| \\
%         =&\frac{1}{N_t}|\sum_{n=-\frac{N}{2}}^{\frac{N}{2}}\text{exp}(j\pi\overline{\gamma} n)| \\
%         =&\frac{1}{N_t}|\int_{-\frac{N}{2}}^{\frac{N}{2}}\text{exp}(j\pi\overline{\gamma} n)\text{d}n| \\
%         =&|\frac{\sin(\frac{N_t\overline{\gamma}}{2})}{N_t\sin(\frac{\overline{\gamma}}{2})}|
%     \end{align}
% \end{subequations}
 where $\overline{\gamma}=\frac{2\pi}{\lambda}(\sin{\theta_n}-\sin{\theta_m})d$.
 %When $\overline{\gamma}=\frac{2u\pi}{N_t}$, where $u=1,2,\cdots,N_t$, $C_{\theta_m,\theta_n}=0$.
When $C_{\theta_m,\theta_n}$ is 0, the corresponding $\overline{\gamma}$ satisfies $\overline{\gamma} = \frac{2\pi u}{N_t}$, where $u = 1, 2, \cdots, N_t$, which can be used to design orthogonal angular intervals.

 Since we obtain the expression of $C_{a_m,a_n}(\overline{\alpha})$, $C_{r_m,r_n}(\overline{\beta})$ and $C_{\theta_m,\theta_n}(\overline{\gamma})$, we can select the appropriate correlation coefficients along with their corresponding sampling intervals for $a$, $r$ and $\theta$ to design the codebooks for beam search. The details of this process are provided in the next subsection.
 %$\Delta\sin\theta=\frac{2u}{N_t},u=1,\cdots,N$
 %We denote $E(\overline{\gamma})=\frac{\sin(\frac{N_t\overline{\gamma}}{2})}{N_t\sin(\frac{\overline{\gamma}}{2})}$
% %(24a)可以转化为
% \begin{subequations}
% \begin{align}
%     C\approx&|C_{a_m,a_n}C_{r_m,r_n}C_{\theta_m,\theta_n}|\\
%     =&|\frac{A(\overline{\alpha})}{\alpha}\frac{B(\overline{\beta})+jD(\overline{\beta})}{\overline{\beta}}E(\overline{\gamma})|
% \end{align}
% \end{subequations}

% \begin{subequations}
%     \begin{align}
%            C&\approx \frac{1}{N_t}|\int_{-\frac{N}{2}}^{\frac{N}{2}}e^{jkH(n)}e^{jkI(n)}e^{jkJ(n)} \text{d}n|\\
%         &\leq \frac{1}{N_t}|\int_{-\frac{N}{2}}^{\frac{N}{2}}e^{jkH(n)} \text{d}n|\\
%     %     &=\frac{1}{N_t}\int_{-\frac{N}{2}}^{\frac{N}{2}}|e^{jkH(n)}||e^{jkI(n)}||e^{jkJ(n)}| \text{d}n\\
%     %        &\leq \frac{1}{N_t}\int_{-\frac{N}{2}}^{\frac{N}{2}}\text{exp}( jk((a_m-a_n)d^3n^3\nonumber\\&+(\frac{\cos^2{\theta_m}}{2r_m}-\frac{\cos^2{\theta_n}}{2r_n})d^2n^2
%     % +(\sin{\theta_n}-\sin{\theta_m})dn) \text{d}n|
%     \end{align}
%     denote $H(n)=(a_m-a_n)d^3n^3$, $I(n)=(\frac{\cos^2{\theta_m}}{2r_m}-\frac{\cos^2{\theta_n}}{2r_n})d^2n^2$ and $J(n)=(\sin{\theta_n}-\sin{\theta_m})dn$
% \end{subequations}

\begin{figure*}[ht]
\centering
    \subfigure[Exhaustive beam search]{\includegraphics[width=0.30\linewidth]{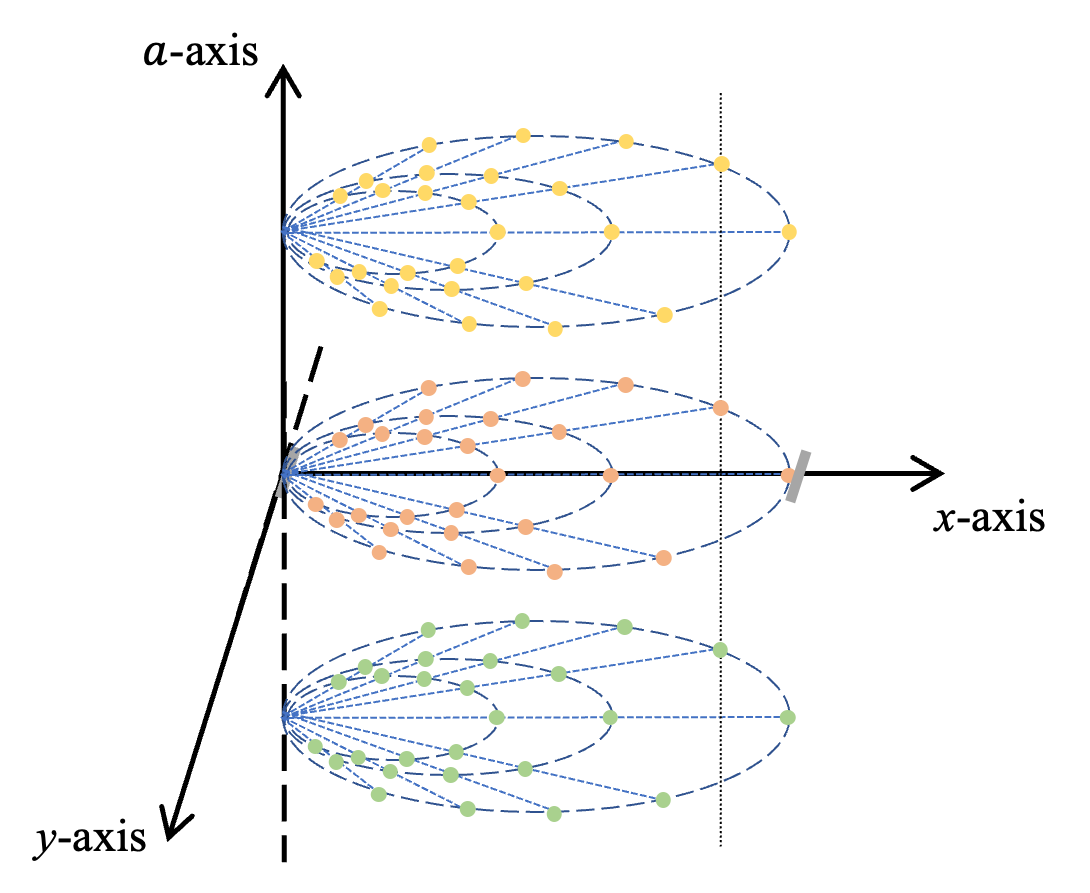}\label{fig:exhaustive_search}} 
     \subfigure[Hierarchical focusing-Airy beam search]{\includegraphics[width=0.30\linewidth]{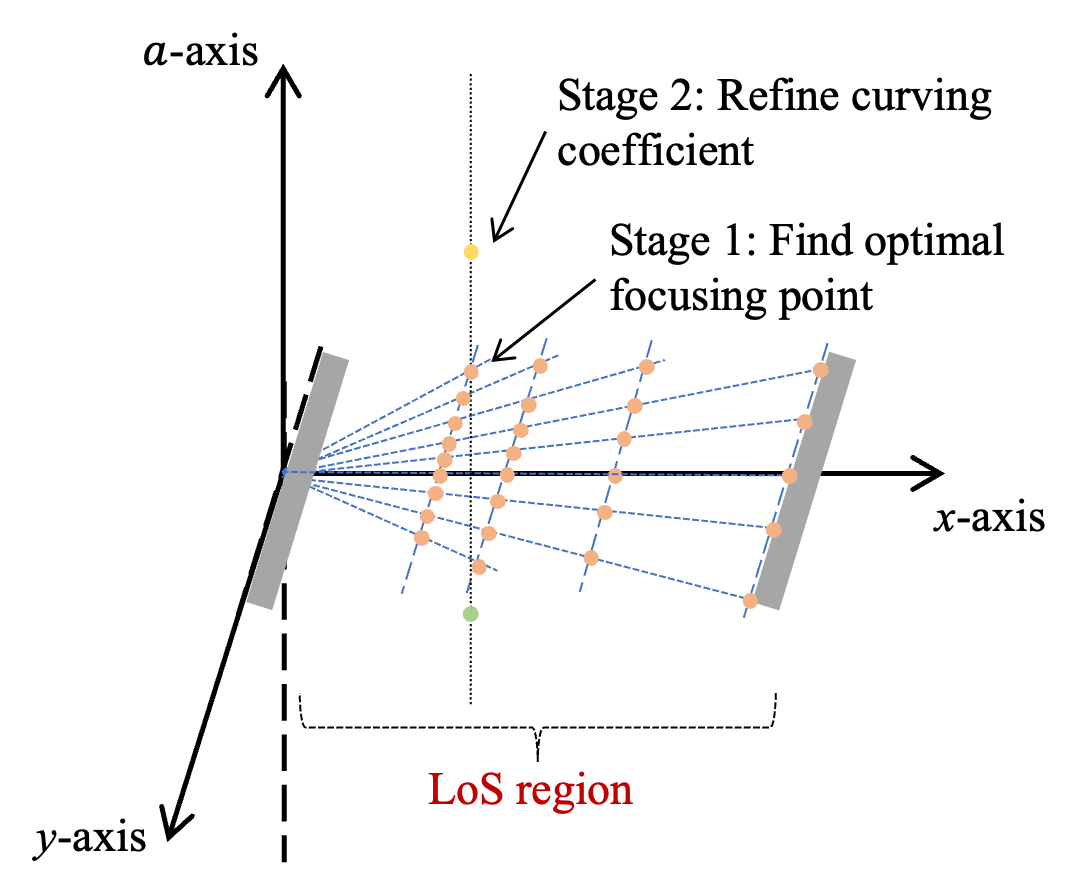}   \label{fig:focusing_Airy_beam_search}}
     \subfigure[Low-complexity beam search]{\includegraphics[width=0.30\linewidth]{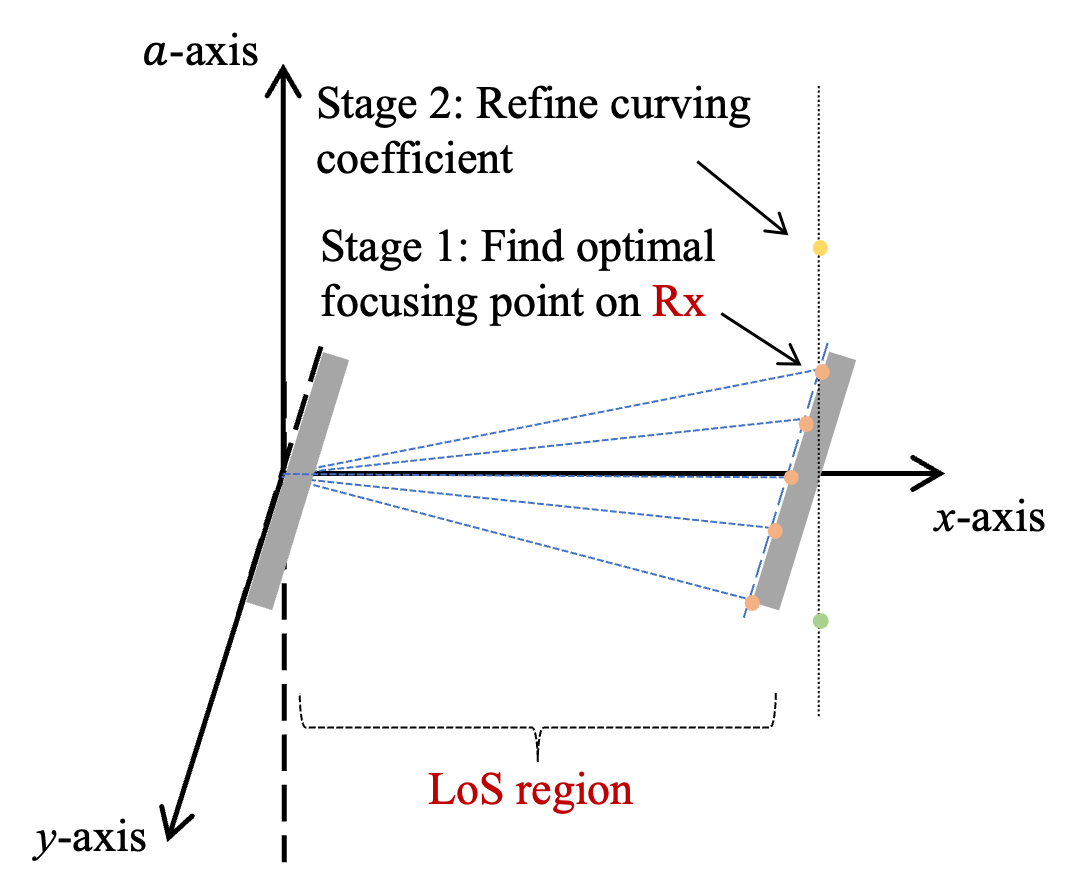} \label{fig:low_complexity_beam_search}}
\caption{Comparison of beam search spaces under different Airy beam search schemes: (a) exhaustive search over all spatial directions and curving coefficients, (b) hierarchical beam search with focusing and curving refinement and (c) low-complexity search based on Rx position information.}
\end{figure*}

\subsection{Beam Sampling Interval}
For Airy beam search, an optimal beam vector is selected from a pre-designed beam codebook through a training procedure between the Tx and Rx. In conventional far field beam searching and near field beam focusing searching, the codewords in their codebook are determined by a unique spatial angle or point location. Similarly, for the Airy beam codebook, we aim to identify a distinctive feature that can uniquely determine an Airy beam codeword. Specifically, since the Airy beam can only exist in near field region, as the analysis in Sec. \ref{Sec:Analysis of Airy}, the Airy beam can be defined through a point location information due to its focusing property and curving coefficient due to its self-accelerating trajectory. Thus, an Airy beam codeword can be uniquely defined by a tuple of parameters, i.e., $(a,r,\theta)$. Each codeword determines a unique three-dimensional point location, specified by the two-dimensional focusing point location and the curving coefficient. The codebook is designed to cover the desired range of these parameters, ensuring comprehensive representation of the target Airy beams within the specified region.

% As shown in Fig. \ref{fig:exhaustive_search}, for the exhaustive beam search, the codebook should contain all the points determined by $(a,r,\theta)$ in three-dimensional space. Since it is infeasible to exhaustively traverse all points in three-dimensional space, we instead sample the 3D space based on the analysis presented in the previous section. 
Based on the beam correlation analysis, the beam sample interval can be derived to design the codebook. Specifically, we assign the values $C_{a_m,a_n}(\overline{\alpha})$, $C_{r_m,r_n}(\overline{\beta})$ and $C_{\theta_m,\theta_n}(\overline{\gamma})$ as $\xi_a$, $\xi_r$ and $\xi_\theta$, respectively, and derive the corresponding $\overline{\alpha}_\xi$,  $\overline{\beta}_\xi$ and  $\overline{\gamma}_\xi$. Subsequently, the sample intervals can be expressed as 
%最后所有的公式都要整理成\left \right的形式
\begin{subequations}
\label{eq:sample}
\begin{align}
   s_a&=|a_m-a_n|=\frac{\overline{\alpha}_\xi^3}{d^2N_t^3}\\
   s_r&=|\frac{\cos^2{\theta}}{r_m}-\frac{\cos^2{\theta}}{r_n}|=\frac{\overline{\beta}_\xi^2}{dN_t^2}\label{eq:r_sample}\\
   s_\theta&=|\sin{\theta_m}-\sin{\theta_n}|=\frac{\overline{\gamma}_\xi}{\pi}=\frac{2u}{N_t}\label{eq:theta_sample},
\end{align}
\end{subequations}
where in \eqref{eq:r_sample}, $\theta$ is a fixed parameter and in \eqref{eq:theta_sample} $u=1,2,\cdots,N_t$. We define $[a_\mathrm{min},a_\mathrm{max}]$ as the potential range of the curving coefficient. Specifically, since the location of the blockage is unknown, the possibility of concave and convex trajectory should be covered in $[a_\mathrm{min},a_\mathrm{max}]$. Thus, we let $ a_\mathrm{min}<0 $, $ a_\mathrm{max}>0 $ and $|a_\mathrm{min}|=|a_\mathrm{max}|$. We denote $J$ as the number of sample curving coefficient and the $j^{\mathrm{th}}$ sample value can be expressed as 
\begin{equation}
    a_j=a_\mathrm{min}+(j-1)s_a,
\end{equation}
where $j=1,2,\cdots,J$ and $a_\mathrm{max}=a_\mathrm{min}+(J-1)s_a$. Moreover, we define $r_\mathrm{max}$ as the distance between the Tx and Rx, i.e., $r_\mathrm{max}=D$. Therefore, in the fixed angle $\theta$, $\frac{\cos^2{\theta}}{r_\mathrm{max}}$ is lowest. We denote $K$ as the number of sample distance and the $k^{\mathrm{th}}$ sample distance can be derived as
\begin{equation}
    r_k=\frac{\cos^2{\theta}}{\frac{\cos^2\theta}{r_\mathrm{max}}+(k-1)s_r},
\end{equation}
where $k=1,2,\cdots,K$. For the angle sample, we define $\theta_\mathrm{min}$ and $\theta_\mathrm{max}$ as $-\frac{\pi}{2}$ and $\frac{\pi}{2}$, respectively. We denote $V$ as the number of sample angle and the $v^{\mathrm{th}}$ sample angle can be derived as
\begin{equation}
    \theta_v=\arcsin(\sin{\theta_\mathrm{min}}+(v-1)\frac{2u}{N_t}),
\end{equation}
where $v=1,2,\cdots,V$ and $u$ determines the sample interval of angle domain. As shown in Fig. \ref{fig:exhaustive_search}, for the exhaustive beam search, the codebook should contain all the sample points determined by $(a,r,\theta)$ in three-dimensional space. Therefore, the total number of codewords in the exhaustive beam search codebook is given by $JKV$, and the corresponding codebook can be expressed as
\begin{equation}
\mathcal{F}_e =\{\mathbf{f}= \mathbf{a}(a_j, r_k, \theta_v)\mid a,r,\theta \in [A,R,\Theta]\}.
\end{equation}
where $A,R,\Theta$ are the set of three parameters, respectively. 

In the Tx, a codeword is selected from the codebook for transmission in each time slot. The Rx employs an omnidirectional analog combiner and a unit matrix digital combiner to process the received signal. The received signal in $t^{\mathrm{th}}$ time slot with $t^{\mathrm{th}}$ codeword $f_t$ can be denoted as
\begin{equation}
    \mathbf{y}_{t}=\mathbf{W}^H(\mathbf{H}_b\mathbf{f}_{t}+\mathbf{n)},
\end{equation}
where $\mathbf{W}$ is combiner in the Rx and $\mathbf{H}_b$ is the blocked channel. Since the number of codewords in the exhaustive codebook is $JKV$, the total number of time slots is $T_e=JKV$. After $T_e$ time slots, the receiver can select an optimal beam which can be described as 
\begin{equation}
    \mathbf{f}_\mathrm{opt}=\arg \max_{\mathbf{f}_\mathrm{t}}||\mathbf{y}_t||^2.
\end{equation}
Finally, an Airy beam vector $\mathbf{f}_\mathrm{e}=\mathbf{a}(a_\mathrm{e},r_\mathrm{e},\theta_\mathrm{e})$ can be used to communicate with the Rx in the quasi-LoS scenario.

\subsection{Hierarchical Focusing-Airy Beam Search Scheme}
To reduce the overhead of exhaustive beam search, we propose a hierarchical Focusing-Airy beam search scheme that narrows the spatial search space and performs a two-stage refinement, where the first stage searches over focusing beams to determine distance and angle and the second stage refines the curving coefficient.
%Although the exhaustive beam search codebook can cover almost all the Airy beams in space, the large number of codewords results in significant overhead for beam search. To reduce the overhead, we continuously propose a hierarchical beam search scheme named focusing-Airy beam search. 
As shown in Fig. \ref{fig:focusing_Airy_beam_search}, firstly, we narrow the scope of the beam search in the spatial domain to reduce the number of codewords. Since the LoS is blocked, the beam search can be initiated in the LoS region within the spatial domain between the Tx and Rx. Suppose that the numbers of antennas in the Tx and Rx are the same and the length of the ULA is $L_a$. The polar coordinate of the points located in the LoS range can be expressed as %记得在图中补充Tx 和 Rx的长度示意
\begin{align}
    \text{point}_\mathrm{LoS}=\{(r,\theta)\mid 0\leq r\cos{\theta}\leq D,\nonumber\\-\frac{L_a}{2}\leq  r\sin{\theta}\leq\frac{L_a}{2}, r\in R, \theta\in \Theta \}.
\end{align}
The number of $\text{point}_\mathrm{LoS}$ is denoted as $T_f^1$. After we determine the scope of beam search in the spatial domain, we propose a two-stage beam search scheme. In the first stage, we set $a$ equal to 0, which means that all the beams in the codebook of the first stage are focusing beams. The codebook for the first stage can be denoted as 
\begin{equation}
    \mathcal{F}_f^1=\{\mathbf{f}=\mathbf{a}(0,r,\theta)\mid (r,\theta)\in \text{point}_\mathrm{LoS}\}.
\end{equation}
After $T_f^1$ time slots, the optimal focusing beam $\mathbf{f}_f=\mathbf{a}(0,r_f,\theta_f)$ can be selected from $\mathcal{F}_f$. In the second stage, we fix the distance coefficient $r_f$ and angle coefficient $\theta_f$ and design a codebook $\mathcal{F}_{a}$ to search the Airy beam with different curving coefficient, which can be expressed as
\begin{equation}
    \mathcal{F}_f^2=\{\mathbf{f}=\mathbf{a}(a,r_f,\theta_f)\mid a\in A\}.
\end{equation}
After $T_f^2$ time slots where $T_f^2$ is the total number of codewords in $\mathcal{F}_f^2$, a near-optimal Airy beam vector can be obtained to keep good performance in quasi-LoS communication scenario. 

Instead of exhaustively searching through all possible Airy beams in the scenario, the hierarchical focusing-Airy beam search scheme reduces the search space by narrowing down the potential beam directions and conducting the hierarchical search, which significantly reduces overhead and speeds up the search, making it more practical for real-time applications.
%线性规划，保证focusing point 在 LoS range内，这里应该考虑发射端接收端不一样天线数量，然后？？？？
\subsection{Low-complexity Beam Search Scheme}
%To further reduce the overhead and accelerate the beam search process, we propose a low complexity beam search for fast and reliable communication establishment in quasi-LoS scenario. 
To enable fast and reliable link recovery in quasi-LoS scenarios, we propose a low-complexity beam search scheme that leverages prior Rx position knowledge to directly focus the beam on the Rx and refine the curving coefficient with significantly reduced overhead. Specifically, as shown in Fig.~\ref{fig:low_complexity_beam_search}, We choose the focusing point on the curve $\frac{\cos{\theta}}{r}=\frac{1}{D}$. Similar to the hierarchical focusing-Airy beam search, we also choose the point located in the LoS range. We define the angle range$[\theta_l^{min},\theta_l^{max}]$ where $\theta^{max}_l$ and $\theta^{min}_l$ are set as $\arctan{\frac{L_a}{2D}}$ and $-\arctan{\frac{L_a}{2D}}$. We set the curving coefficient $a$ as 0, the angle sample interval as $s_\theta$ and design a codebook $\mathcal{F}_{l}^1$ for searching an appropriate focusing beam, which can be denoted as
\begin{equation}
    \mathcal{F}_{l}^1=\{\mathbf{f}=\mathbf{a}(0,r,\theta)\mid \frac{\cos{\theta}}{r}=D,\theta\in[\theta^l_{min},\theta^l_{max}]\}.
\end{equation}
After $T_l^1$ time slots where $T_l^1$ is the number of the codewords in $\mathcal{F}_{l}^1$, the selected focusing beam vector can be obtained as $\mathbf{f}=\mathbf{a}(0,r_l,\theta_l)$. Then we fix the $r_l$ and $\theta_l$ to search the Airy beam with different curving coefficients, which can be expressed as
\begin{equation}
    \mathcal{F}_l^2=\{\mathbf{a}(a,r_l,\theta_l)\mid a\in A\}.
\end{equation}
After $T_l^2$ time slots where $T_l^2$ is the number of the codewords in $\mathcal{F}_{l}^2$, the sub-optimal beam vector $\mathbf{f}=\mathbf{a}(a_l,r_l,\theta_l)$ can be obtained to fast recover the communication link in quasi-LoS scenarios.

Although the performance of the spectral efficiency using low complexity beam search is worse than that of exhaustive search and hierarchical focusing-Airy search, its overhead is significantly reduced compared to these two schemes. Therefore, the low complexity beam search scheme can be used for fast communication link recovery and reestablishment in the data center when short-term blockages occur, ensuring reliable connectivity for critical tasks that depend on these specific links.
%再加上如何去评估beam searching好坏的方法，也就是计算SNR信道容量
\section{Performance Evaluation}
\label{sec: evaluation}
In this section, simulations are provided to evaluate the accuracy of the proposed CGWCM and the performance of the proposed three Airy beam search schemes. We consider a THz hybrid beamforming system, where the Tx and Rx equip ULA with 256 antennas. The carrier frequency is $f_c=140$~GHz. The channel in the data center is generated by Quadriga based on the measurement data in \cite{Channel-2022-Song}.
%后续在讲codebook的preformance的时候，讲一下是如何生成被遮挡信道的，因为【】中测量的情况是不包含遮挡信道的，所以，需要去讲LoS用提出的信道模型基于没有被遮挡的信道重新做了调整，非视距信道的数据保留。

%图片中CGWCM的红色换成比较鲜艳一点的红色。
\subsection{Accuracy of CGWCM Channel Model}
\begin{figure*}[ht]
    \centering
    \subfigure[Approximation error versus height of blockage.]{\includegraphics[width=0.32\linewidth]{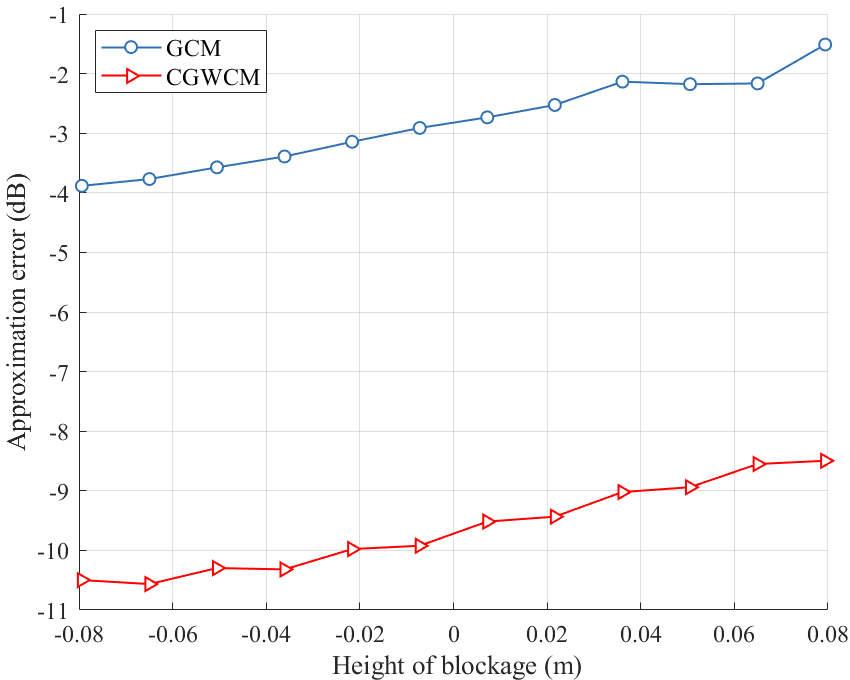}\label{fig:Approxi_error_CM}} 
     \subfigure[The power of the received signal in the Rx.]{\includegraphics[width=0.32\linewidth]{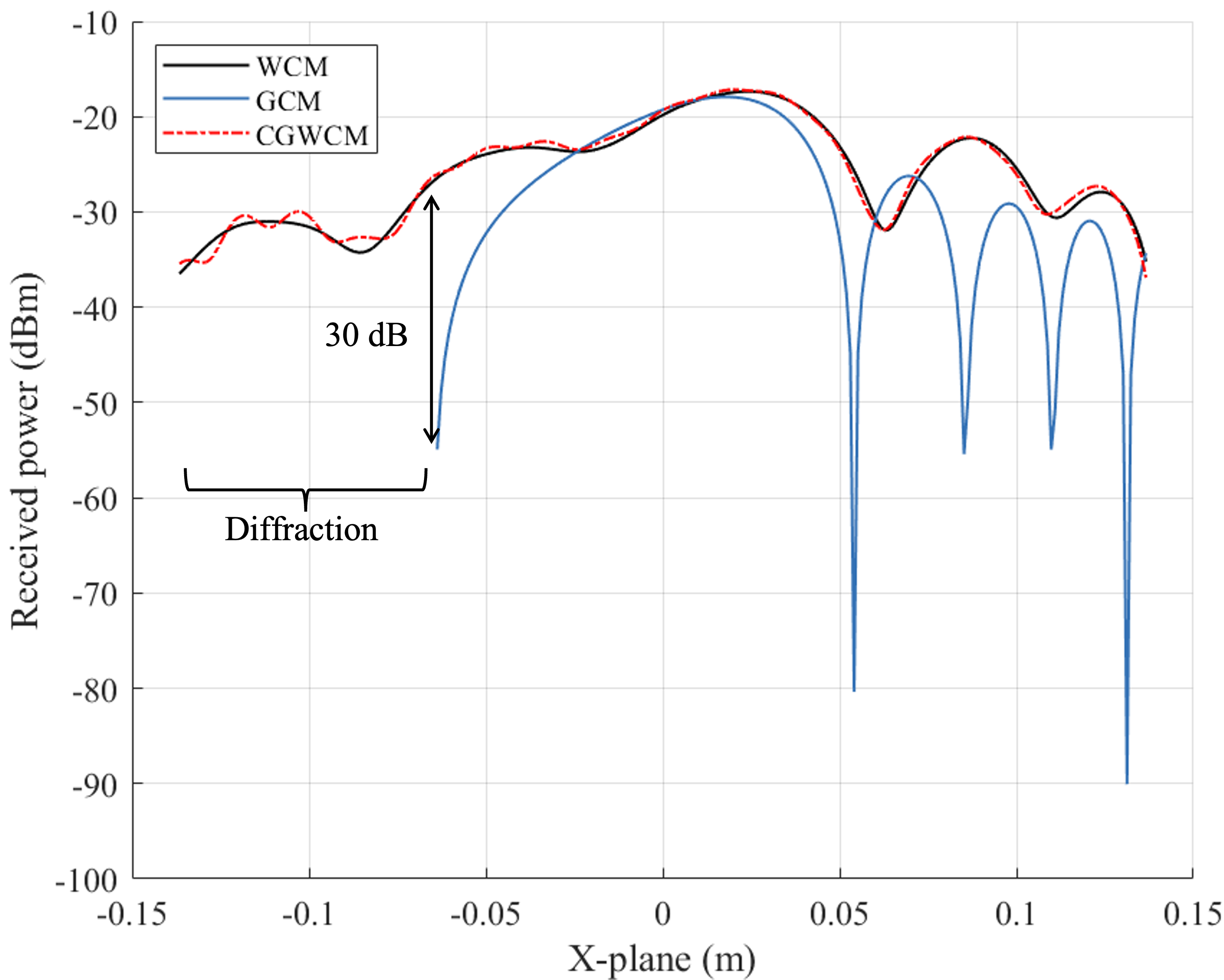}\label{fig:gain_error_CM}}
     \subfigure[The phase of the received signal in the Rx. ]{\includegraphics[width=0.32\linewidth]{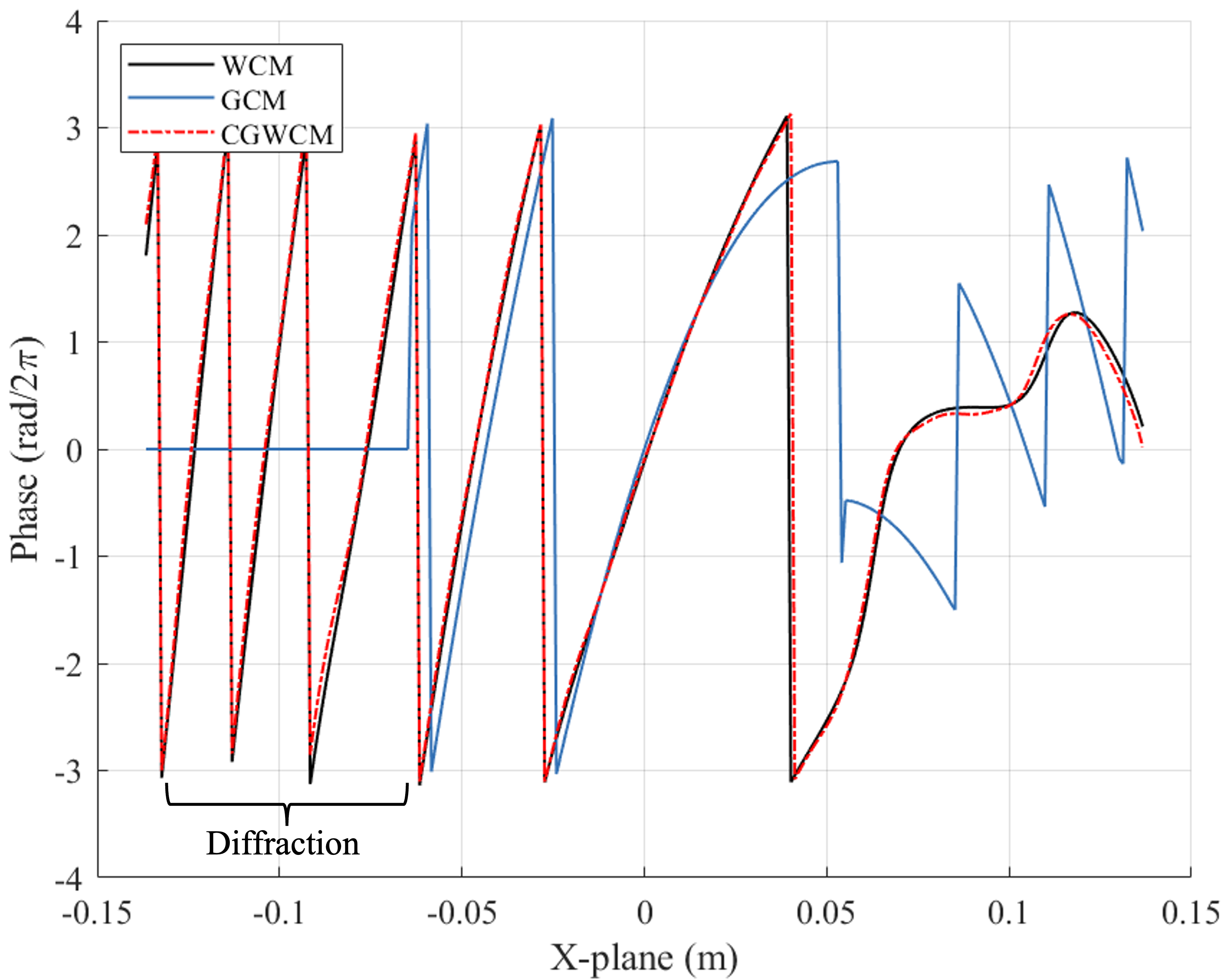}\label{fig:phase_error_CM}}
   \caption{The accuracy of the GCM and CGWCM.}
   \label{fig:CM_accuracy}
\end{figure*}

We begin by evaluating the accuracy of the proposed CGWCM in Fig.~\ref{fig:Approxi_error_CM}, where the approximation errors of both GCM and CGWCM are calculated under varying blockage heights ranging from [-0.08, 0.08 m]. The approximation errors of the GCM and CGWCM are obtained by calculating $||\mathbf{H}_\mathrm{G}-\mathbf{H}_\mathrm{W}||_\mathrm{F}/||\mathbf{H}_{\mathrm{W}}||_\mathrm{F}$ and $||\mathbf{H}_\mathrm{C}-\mathbf{H}_\mathrm{W}||_\mathrm{F}/||\mathbf{H}_{\mathrm{W}}||_\mathrm{F}$, respectively. First, the CGWCM remains highly accurate in the quasi-LoS scenario, in which the approximation error of the CGWCM is much smaller than that of the GCM under different heights of blockage. This is because the diffraction is explored in the CGWCM, while GCM only considers the straight propagation of the EM wave. Specifically, as shown in Fig. \ref{fig:Approxi_error_CM}, with $D=3$ m and $L=1.5$ m, the approximation error of the CGWCM is 7~dB lower than the GCM when the height of the blockage is close to 0~m. Moreover, the approximation error of the GCM increases as the blockage height rises, reaching -2~dB when the entire LoS path is nearly fully obstructed. In contrast, the CGWCM maintains a significantly lower error of around –8.5~dB. This illustrates that in complex blockage scenarios, the CGWCM maintains better accuracy than the conventional GCM.

As shown in Fig. \ref{fig:gain_error_CM} and Fig. \ref{fig:phase_error_CM}, the power and phase of the signal in each antenna in the Rx are explored with fixed height of the blockage as 0.036 m. Specifically, we set the analog precoder in the Tx with the focusing beam phase vector $\mathbf{a}(a=0,r=D,\theta=0)$ then the power and phase distribution on the Rx antenna array with different channel models by calculating abs($\mathbf{H}_{\mathrm{(G,W,C)}}\mathbf{F}_{\mathrm{RF}}$) and angle($\mathbf{H}_{\mathrm{(G,W,C)}}\mathbf{F}_{\mathrm{RF}}$), respectively. In terms of both power and phase, the CGWCM closely matches the WCM, whereas the GCM exhibits significant difference from the WCM. Furthermore, as shown in Fig. \ref{fig:gain_error_CM}, focusing on the blocked region ranging from -0.136~m to -0.063~m, the received power calculated based on the WCM and CGWCM is substantially higher than that based on the GCM. This indicates that even when the direct path between the Tx and Rx is completely blocked in this region, diffraction allows a portion of the signal power to reach these antennas, demonstrating that signal propagation follows a curved rather than a strictly straight trajectory. This phenomenon, which cannot be captured by the GCM, is effectively characterized by both the WCM and CGWCM.
\subsection{Performance of Codebook Design}
% 不同beam的比较，横坐标发射功率，纵坐标SE这样
% 障碍物高度变化
% 障碍物距离发射端距离变化
% 接收端天线数量变化
% 多数据流的处理 （在讲codebook的时候，后面最后说流程的时候也要去说不同data streams）
In this subsection, the performance of the proposed Airy beam search schemes are evaluated. The distance between the Tx and Rx is 3 m and the blockage position varies within $[0, 3 \text{ m}]$ horizontally and $[-0.136, 0.136 \text{ m}]$ vertically. The non-blocked channel is generated by Quadriga consisting of a LoS path and multiple NLoS paths. According to the CGWCM in Sec.~\ref{sec:CGWCM}, the quasi-LoS channel matrix can be modeled given the position of the blockage. We consider that the blockage only influences the LoS path, then the blocked channel can be obtained by replacing the LoS path in the non-blocked channel with the calibrated quasi-LoS channel. For the proposed Airy beam search schemes, according to the analysis in \eqref{eq:sample}, we assign the values $\xi_a=0.4$, $\xi_r=0.15$ and $\xi_\theta=0$ and derive the corresponding $\overline{\alpha}_\xi\approx1.69$, $\overline{\beta}_\xi\approx4.59$ and $\overline{\gamma}_\xi\approx0.0245$. The sample intervals are $s_a\approx0.25$, $s_r\approx\frac{1}{3}$ and $s_\theta\approx0.0078$. To quantify the performance of the beam search schemes, we calculate the spectral efficiency as
\begin{equation}
    R=\log_2(|\mathbf{I}_{N_s}+\frac{\rho}{N_s}\mathbf{R}_n^{-1}\mathbf{W}^H\mathbf{H}_b\mathbf{F}\mathbf{F}^H\mathbf{H}_b^H\mathbf{W}|),
\end{equation}
where $\mathbf{F}$, $\mathbf{W}$ and $\mathbf{R}_n=\sigma^2\mathbf{W}^H\mathbf{W}$ are precoder, combiner and noise covariance matrix, respectively.

To assess the proposed Airy beam search schemes, we set several benchmarks, which are shown below.
\begin{itemize}
\item[$\bullet$] \textbf{Non-blocked precoder (LoS)}: The full digital precoder is designed base on LoS channel, and the spectral efficiency is calculated under the non-blocked scenario.
\item[$\bullet$] \textbf{Non-blocked precoder (quasi-LoS)}: The same precoder from the LoS case is used, but spectral efficiency is evaluated under the blocked scenario.
\item[$\bullet$] \textbf{Perfect CSI}: Assuming perfect knowledge of the blocked channel $\mathbf{H}_b$, a full-digital precoder is designed to provide an upper bound on spectral efficiency in the blocked scenario.
\item[$\bullet$] \textbf{Far-field steering (Gaussian)}\cite{Alkhateeb-2015-Limited}: This method is a classical far-field beam steering which searches the optimal beam only in the angle domain.
\item[$\bullet$] \textbf{Near-field focusing (Gaussian)}\cite{Zhang-2022-Beam-focusing}: This method is a near-field beam focusing which searches the optimal beam focusing on the antenna of the Rx array.
\item[$\bullet$] \textbf{NLoS}: Based on perfect knowledge of the NLoS channel $\mathbf{H}_{\text{NLoS}}$ generated using Quadriga \cite{Channel-2022-Song}, a full-digital precoder is designed to evaluate the spectral efficiency using only NLoS paths.
\end{itemize}
It is worth noting that when the LoS becomes quasi-LoS, we aim to avoid additional overhead and resource consumption for repeatedly estimating the blocked channel information. In other words, after using a few pilots to search for the optimal beams, we design the precoder based on the analysis in \eqref{eq:precoder} with a non-blocked channel, which is obtained during the initial establishment of the communication link. Nevertheless, the spectral efficiency is evaluated using the blocked channel. As verified in the following content, the performance difference between the precoder designed with the non-blocked channel and the one designed with the blocked channel is minimal when using Airy beam for the analog precoder.
\begin{figure}[t]
    \centering
    \includegraphics[width=0.8\linewidth]{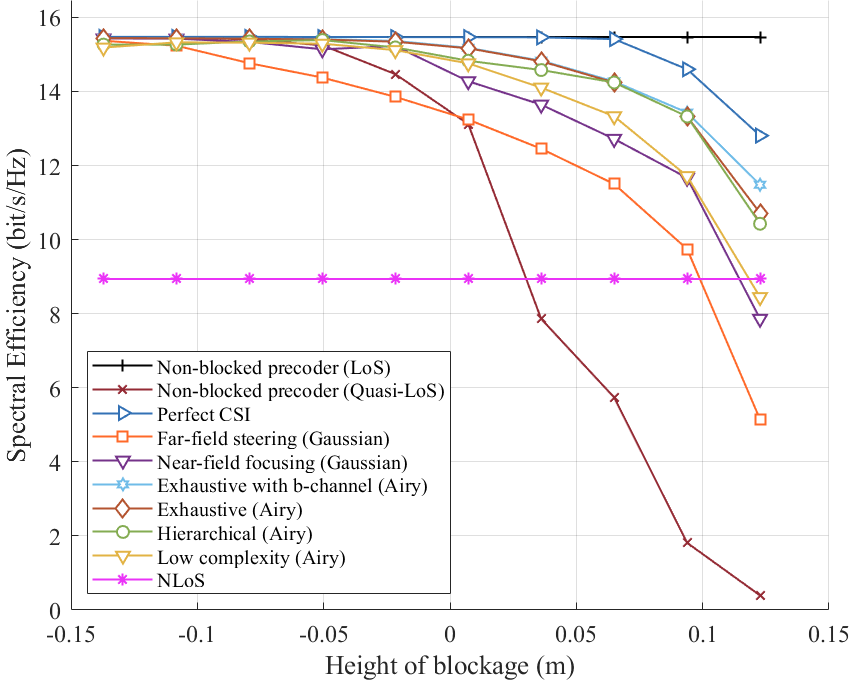} % 使用 \linewidth 自适应栏宽
    \caption{Spectral efficiency versus height of blockage. $D=3$ m,$L=1.5$ m.}
     \label{fig:SE_h}
\end{figure}

First of all, we begin by evaluating the spectral efficiency of all the beam search methods under different heights of the blockage. For analytical simplicity, we evaluate different beam patterns under the assumption of $N_s = 1$. This assumption holds in data centers, where each antenna array on a rack communicates with multiple racks, and in most cases, each link between any two racks corresponds to a single data stream. Fig. \ref{fig:SE_h} illustrates the spectral efficiency versus the height of the blockage. It can be observed that as the height of the blockage increases, the spectral efficiency of most beam search schemes gradually decreases, except for the NLoS link, which remains constant at approximately 9~bits/s/Hz due to the no influence from the blockage. The non-blocked precoder (LoS) holds the highest spectral efficiency, staying above 15.4~bits/s/Hz. In contrast, the non-blocked precoder (quasi-LoS) exhibits a rapid decline, dropping from 15.4~bits/s/Hz at -0.136 m to below 6~bits/s/Hz when the blockage height exceeds 0.05 m. Therefore new beam search schemes are needed to enhance the robustness of the communication system.

For the Airy beam search schemes, exhaustive search achieves the highest spectral efficiency among all the beam search schemes and lower than the perfect CSI situation. Specifically, the exhaustive search with the non-blocked channel have almost the same performance as that with the blocked channel when the height of the blockage is lower than 0.1~m and have 0.7~bits/s/Hz loss at 0.122 m. This demonstrates that it is good enough to design the precoder based on the initially known non-blocked channel and no need for the blocked channel information. However, exhaustive search requires extensive overhead leading to high computational complexity. The Hierarchical method balances the performance and overhead. At 0.05 m, it achieves 14.3~bits/s/Hz, while reducing the computational overhead. The low-complexity method provides a practical alternative with minimal overhead. At 0.05 m, its spectral efficiency is 13.7~bits/s/Hz, which is 0.6~bits/s/Hz and 0.8~bits/s/Hz lower than the hierarchical method and exhaustive method, respectively.

When comparing Airy beam-based and Gaussian-based methods, the superiority of the former becomes evident, particularly under increasing blockage conditions. Specifically, when the height of the blockage exceeds 0 m, i.e. over 50\% of the LoS region being blocked, the average spectral efficiency difference between the far-field method and the exhaustive search, hierarchical search and low-complexity search is approximately 3.23~bits/s/Hz, 3.06~bits/s/Hz and 2.05~bits/s/Hz, respectively. Similarly, the near-field method exhibits an average spectral efficiency deficit of 1.62~bits/s/Hz, 1.45~bits/s/Hz and 0.44~bits/s/Hz compared to the exhaustive search, hierarchical search and low-complexity search.

\begin{figure}[!t]
    \centering
    \subfigure[LoS]{\includegraphics[width=0.8\linewidth]{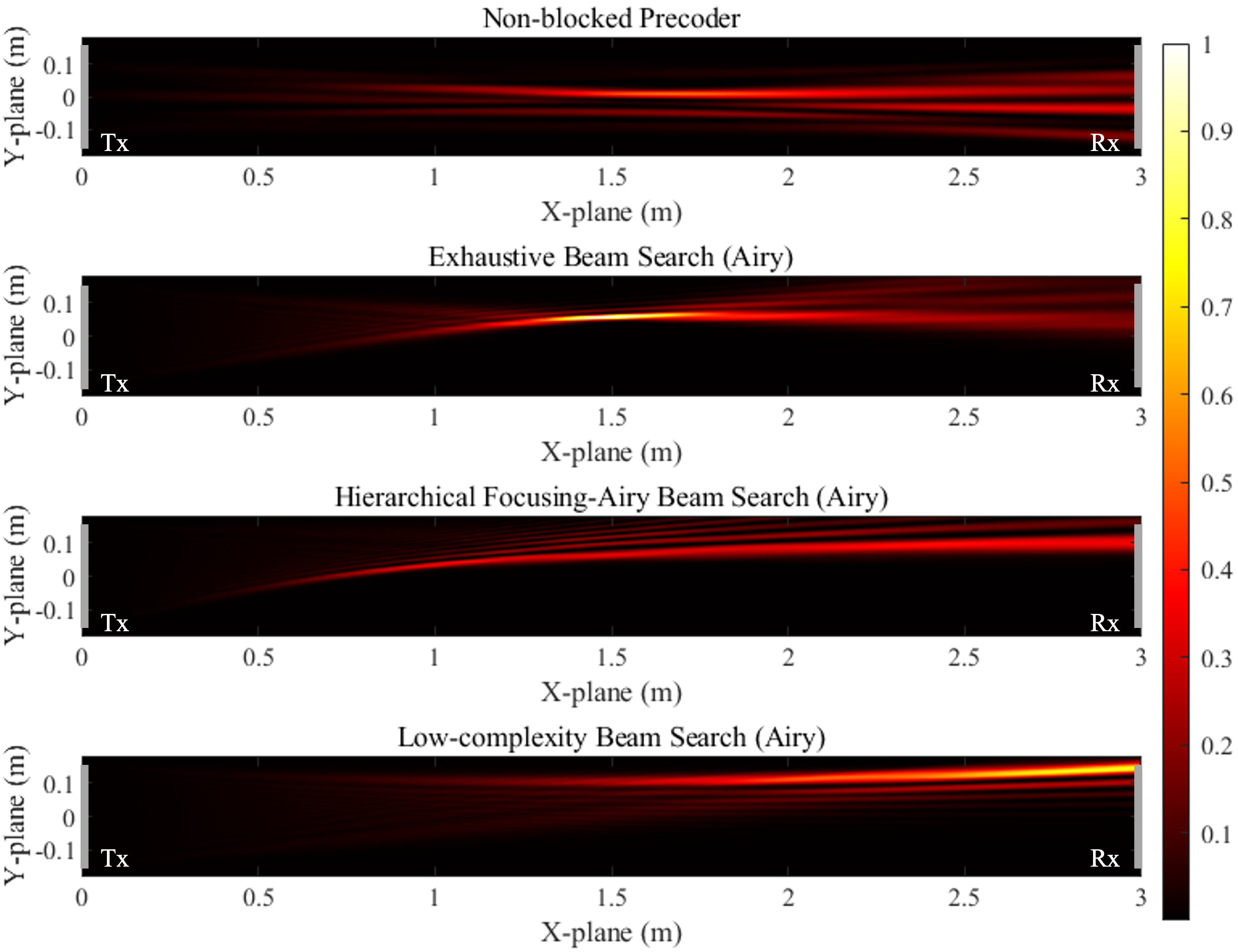}\label{fig:LoS_beam_pattern}} 
     \subfigure[Quasi-LoS]{\includegraphics[width=0.8\linewidth]{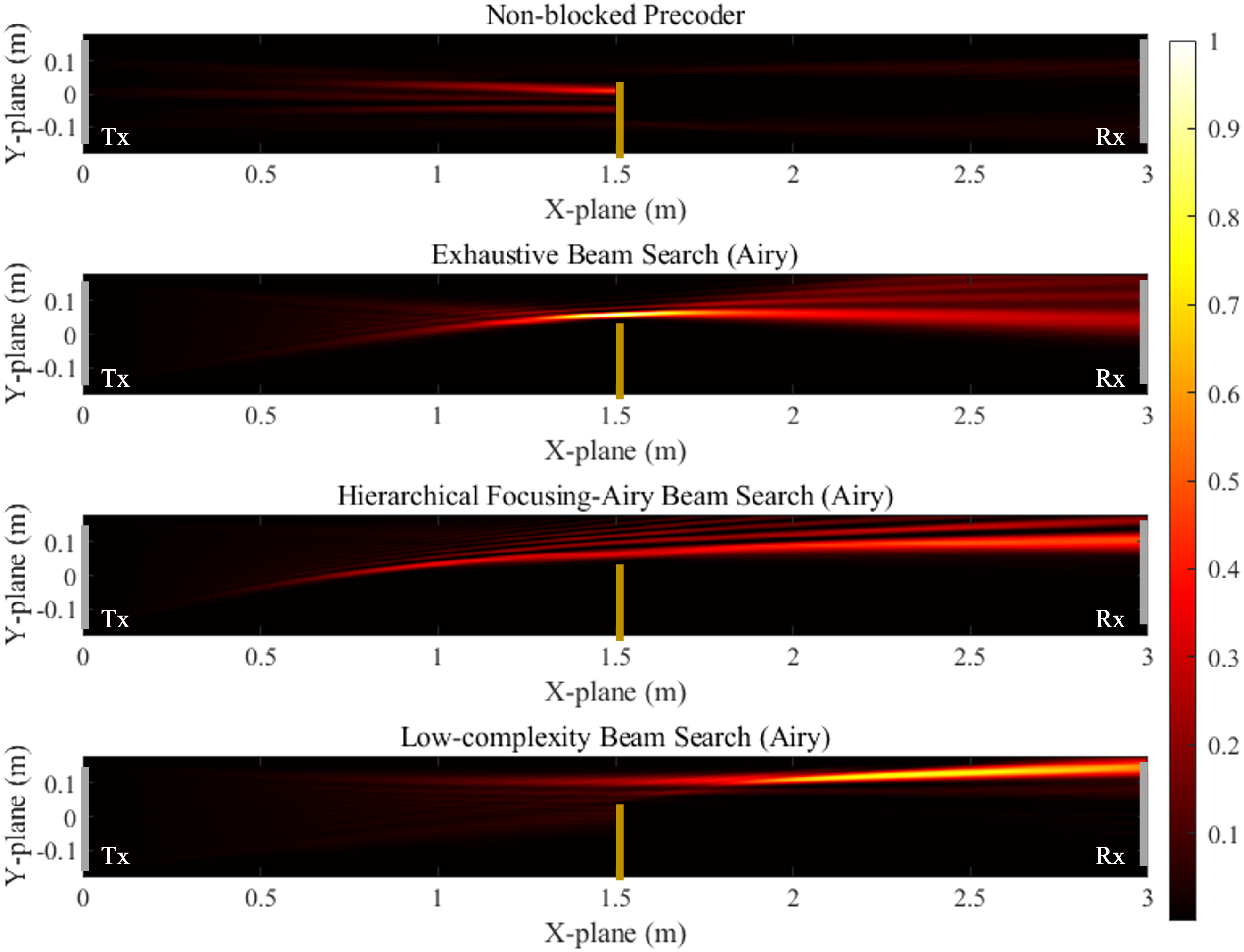}\label{fig:qLoS_beam_pattern}}
    \caption{Beam comparison with different search scheme. $D=3~\textrm{m}, L=1.5~\textrm{m}$ and the height of the blockage is 0.036 m}
\end{figure}
To gain insights of the Airy beam pattern of different beam search schemes, we plot the non-blocked precoder and Airy beams in both LoS and quasi-LoS scenarios. Comparing Fig. \ref{fig:LoS_beam_pattern} and Fig.~\ref{fig:qLoS_beam_pattern}, we observe that the energy of the non-blocked precoder is almost completely blocked. In contrast, the three Airy beams remain unaffected, demonstrating strong robustness against blockage. Specifically, the Airy beams obtained using the exhaustive search and hierarchical search exhibit a convex shape, as if they bend over the blockage. This behavior aligns well with our expectations. The trajectory of the Airy beam obtained from the low complexity is a concave shape, not bending over the blockage, which is different from the exhaustive search and hierarchical search. Similar observations are also obtained in \cite{Chen-2024-Curving} where the signs of the curving coefficients output by neural network (NN) is different from those in exhaustive search indicating that it is not always necessary for the beam to bend over the blockage to establish the communication link. Instead, all Airy beams concentrate energy toward unblocked regions, ensuring that more energy can reach the receiver, thereby achieving good performance.

\begin{figure}[t]
    \centering
    \includegraphics[width=0.7\linewidth]{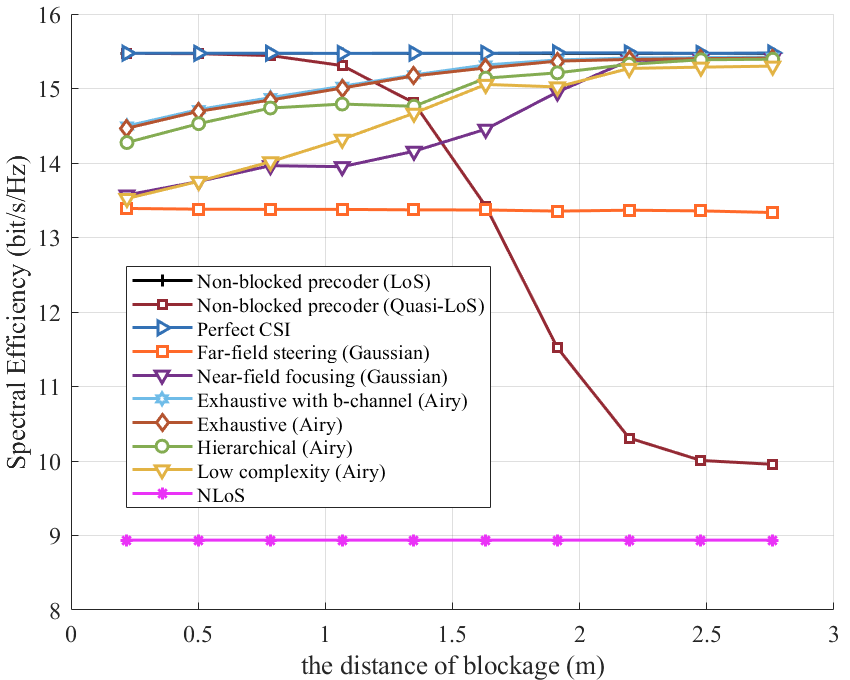} % 使用 \linewidth 自适应栏宽
    \caption{Spectral efficiency versus distance of blockage. D = 3 m and the height of the blockage is 0 m.}
     \label{fig:se_dis}
\end{figure}

As shown in Fig. \ref{fig:se_dis}, we evaluate the spectral efficiency under varying blockage distances. As the blockage distance increases, the performance of the non-blocked precoder (quasi-LoS) deteriorates significantly, whereas the Airy beam search schemes maintain robust performance. Specifically, the hierarchical search method outperforms the far-field and near-field approaches by 1.59~bits/s/Hz and 0.45~bits/s/Hz, respectively. Furthermore, the performance of the low-complexity method improves as the blockage moves closer to the receiver.

% These aforementioned results highlight the robustness of Airy beam-based methods, particularly in mitigating blockage effects, making them more suitable for practical THz communications in wireless data centers.

% \begin{figure}[t]
%     \centering
%     \subfigure[]{\includegraphics[width=0.8\linewidth]{images/simulation_folder/performance/codebook_compare_overload1.png}\label{}} 
%      \subfigure[]{\includegraphics[width=0.8\linewidth]{images/simulation_folder/performance/codebook_compare_overload2.png}\label{}}
%     \caption{Overhead comparison with different searching schemes. D=3 m, L=1.5 m and the height of the blockage is 0.016 m.}
%     \label{fig:SE_overhead}
% \end{figure}
\begin{figure}[t]
    \centering
    \includegraphics[width=0.7\linewidth]{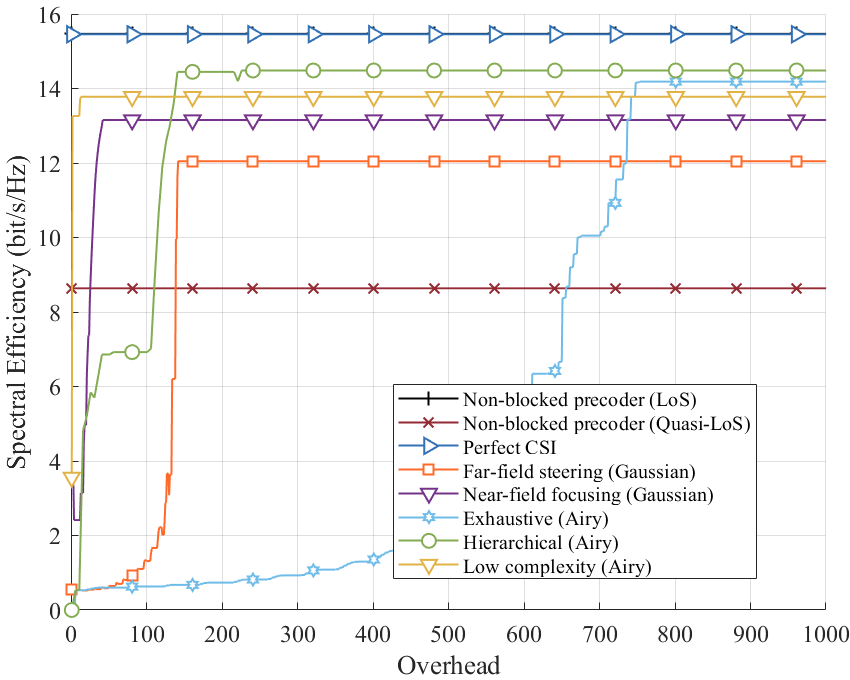} % 使用 \linewidth 自适应栏宽
    \caption{Overhead comparison with different searching schemes. D=3 m, L=1.5 m and the height of the blockage is 0.036 m.}
     \label{fig:SE_overhead}
\end{figure}

Finally, we analyze and compare the beam search overhead among different search schemes. In each time slot, we choose one codeword to transmit and calculate the spectral efficiency utilizing the optimal codeword we search in the previous time slots. As shown in Fig. \ref{fig:SE_overhead}, the exhaustive search method achieves the highest spectral efficiency but at the expense of significantly high overhead. When the overhead reaches 751, the spectral efficiency attains 14.1 bits/s/Hz. However, as calculated in Fig. \ref{fig:SE_h}, the final spectral efficiency converges to 14.8 bits/s/Hz. This indicates that the remaining performance gain requires a substantial increase in overhead, exceeding 1000, which is not depicted in the figure. Second, the hierarchical search method demonstrates a balance between performance and efficiency. As the overhead increases, the spectral efficiency steadily improves, eventually converging to 14.5~bits/s/Hz close to the exhaustive search scheme with only 156 overhead. Notably, hierarchical search achieves near-optimal performance with significantly reduced overhead, making it a promising alternative for practical implementation. The low-complexity method can achieve 13.8 bits/s/Hz, which is a sub-optimal spectral efficiency but requires only 28 overhead. This demonstrates that the low-complexity method is suitable for rapidly reestablishing communication links in the presence of temporary blockage conditions, such as those caused by maintenance personnel. Additionally, far-field and near-field methods (Gaussian) exhibit consistently lower spectral efficiency and more overhead than low-complexity search methods. This shows that the Airy beam-based search schemes are effective in overcoming the blockage and remain robust in quasi-LoS in wireless data centers.

\section{Conclusion}
\label{sec:conclusion}
In this paper, we have investigated the quasi-LoS beamforming problem based on the Airy beam in a near-field THz UM-MIMO system. First, we introduce two fundamental channel models, GCM and WCM, and analytically derived an accurate and low-complexity CGWCM for quasi-LoS scenarios. This model integrates geometrical and wave channel models while transforming the complex integral form into a simple matrix product representation. Extensive simulation results demonstrate that CGWCM achieves higher accuracy than GCM in quasi-LoS scenarios while maintaining a more simplified form compared to WCM.  

Then, we investigate the characteristics of the Airy beam, including its non-diffraction, self-acceleration, and self-healing properties. Furthermore, to generate the Airy beam in the THz UM-MIMO system, we introduce a cubic curving coefficient into the Gaussian phase profile to realize an arbitrary Airy beam generation.
Next, we analyze the beam correlation and derive the sampling interval to the curving, distance and angle coefficients. Based on this, we propose the beam search schemes to establish communication links in quasi-LoS scenarios including hierarchical focusing-Airy beam search, and low-complexity beam search which can find the optimal Airy beam without prior knowledge of blockage position. Extensive simulation results demonstrate that 
with more than 50\% of the LoS region blocked, the hierarchical Airy beam search achieves up to 7.7~bit/s/Hz average improvement over that not considering quasi-LoS, and outperforms the far-field and near-field Gaussian methods by approximately 3.06~bit/s/Hz and 1.45~bit/s/Hz, respectively. Additionally, the low-complexity search maintain performance comparable to the hierarchical search while reducing the overhead by over 80\% . Furthermore, not all Airy beams necessarily bend over the blockage. Instead, they concentrate energy toward unblocked regions, ensuring robust link performance. These results indicate that the Airy beam is more effective than the Gaussian beam in mitigating blockages, making it a promising solution for practical THz communication in wireless data centers.

\bibliographystyle{IEEEtran}
\bibliography{IEEEabrv,Airy_Beam}

\newpage

\end{document}